%% file: dcat.tex
\documentclass[usenatbib]{tex/mn2e}
\usepackage{lscape}
\usepackage{graphicx}
\usepackage{amssymb,amsmath}
\usepackage{natbib}
\usepackage{caption}

\begin{document}

\title{Star Cluster Catalogs for the LEGUS Dwarf Galaxies}

\author[D. Cook et al.]
{D.O. Cook$^{1,2}$, 
J.C. Lee$^{2}$, 
A. Adamo$^{3}$, 
H. Kim$^{4}$, 
R. Chandar$^{5}$, 
B.C. Whitmore$^{6}$, 
\newauthor
A. Mok$^{5}$, 
J.E. Ryon$^{6}$, 
D.A. Dale$^{7}$, 
D. Calzetti$^{8}$,   
J.E. Andrews$^{9}$, 
A. Aloisi$^{6}$, 
\newauthor
G. Ashworth$^{10}$, 
S.N. Bright$^{6}$, 
T.M. Brown$^{6}$, 
C. Christian$^{6}$,
M. Cignoni$^{11}$, 
\newauthor
G.C. Clayton$^{12}$, 
R. da Silva$^{13}$,  
S.E. de Mink$^{14}$, 
C.L. Dobbs$^{15}$, 
B.G. Elmegreen$^{16}$, 
\newauthor
D.M. Elmegreen$^{17}$, 
A.S. Evans$^{18,19}$, 
M. Fumagalli$^{20}$, 
J.S. Gallagher III$^{21}$, 
\newauthor
D.A. Gouliermis$^{22,23}$, 
K. Grasha$^{8}$, 
E.K. Grebel$^{24}$, 
A. Herrero$^{25,26}$, 
D.A. Hunter$^{27}$, 
\newauthor
E.I. Jensen$^{7}$,   
K.E. Johnson$^{18}$, 
L. Kahre$^{28}$, 
R.C. Kennicutt $^{29,30}$,
M.R. Krumholz$^{31}$, 
\newauthor
N.J. Lee$^{7}$, 
D. Lennon$^{32}$, 
S. Linden$^{18}$, 
C. Martin$^{1}$, 
M. Messa$^{3}$, 
P. Nair$^{33}$, 
A. Nota$^{6}$, 
\newauthor
G. \"Ostlin$^{3}$,
R.C. Parziale$^{7}$,
A. Pellerin$^{34}$,
M.W. Regan$^{6}$,  
E. Sabbi$^{6}$, 
E. Sacchi$^{6}$, 
\newauthor
D. Schaerer$^{35}$, 
D. Schiminovich$^{36}$, 
F. Shabani$^{24}$, 
F.A. Slane$^{7}$, 
J. Small$^{7}$, 
\newauthor
C.L. Smith$^{7}$, 
L.J. Smith$^{6}$, 
S. Taibi$^{25}$, 
D.A. Thilker$^{37}$, 
I.C. de la Torre$^{7}$, 
M. Tosi$^{38}$, 
\newauthor
J.A. Turner$^{7}$, 
L. Ubeda$^{6}$
S.D. Van Dyk$^{2}$,  
R.A.M. Walterbos$^{28}$, 
A. Wofford$^{39}$ 
\\
$^1$Department of Physics \& Astronomy, California Institute of Technology, Pasadena, CA 91101, USA; dcook$@$astro.caltech.edu\\ 
$^2$Infrared Processing and Analysis Center, California Institute of Technology, Pasadena, CA\\
$^3$Dept. of Astronomy, The Oskar Klein Centre, Stockholm University, Stockholm, Sweden \\
$^{4}$Gemini Observatory, Casilla 603, La Serena, Chile\\
$^5$Dept. of Physics and Astronomy, University of Toledo, Toledo, OH\\
$^6$Space Telescope Science Institute, Baltimore, MD\\
$^7$Dept. of Physics and Astronomy, University of Wyoming, Laramie, WY\\
$^8$Department of Astronomy, University of Massachusetts, Amherst, MA 01003, USA\\
$^9$Steward Observatory, University of Arizona, 933 North Cherry Avenue, Tucson, AZ 85721\\
$^{10}$Institute for Computational Cosmology and Centre for Extragalactic Astronomy, University of Durham, Durham, UK\\
$^{11}$Department of Physics, University of Pisa, Largo B. Pontecorvo 3, 56127, Pisa, Italy \\
$^{12}$Dept. of Physics and Astronomy, Louisiana State University, Baton Rouge, LA\\
$^{13}$Dept. of Astronomy \& Astrophysics, University of California -- Santa Cruz, Santa Cruz, CA\\
$^{14}$Astronomical Institute Anton Pannekoek, University of Amsterdam, Amsterdam, The Netherlands\\
$^{15}$School of Physics and Astronomy, University of Exeter, Exeter, United Kingdom\\
$^{16}$IBM Research Division, T.J. Watson Research Center, Yorktown Hts., NY\\
$^{17}$Dept. of Physics and Astronomy, Vassar College, Poughkeepsie, NY\\
$^{18}$Dept. of Astronomy, University of Virginia, Charlottesville, VA\\
$^{19}$National Radio Astronomy Observatory, Charlottesville, VA\\
$^{20}$Institute for Computational Cosmology and Centre for Extragalactic Astronomy, Durham University, Durham, UK\\
$^{21}$Dept. of Astronomy, University of Wisconsin--Madison, Madison, WI\\
$^{22}$Zentrum f\"ur Astronomie der Universit\"at Heidelberg, Institut f\"ur Theoretische Astrophysik, Albert-Ueberle-Str.\,2, 69120 Heidelberg, Germany\\
$^{23}$Max Planck Institute for Astronomy,  K\"onigstuhl\,17, 69117 Heidelberg, Germany\\
$^{24}$Astronomisches Rechen-Institut, Zentrum f\"ur Astronomie der Universit\"at Heidelberg, M\"onchhofstr.\ 12--14, 69120 Heidelberg, Germany\\
$^{25}$Instituto de Astrofisica de Canarias, La Laguna, Tenerife, Spain\\
$^{26}$Departamento de Astrofisica, Universidad de La Laguna, Tenerife, Spain\\
$^{27}$Lowell Observatory, Flagstaff, AZ\\
$^{28}$Dept. of Astronomy, New Mexico State University, Las Cruces, NM\\
$^{29}$Institute of Astronomy, University of Cambridge, Cambridge, United Kingdom\\
$^{30}$Dept. of Astronomy, University of Arizona, Tucson, AZ\\
$^{31}$Research School of Astronomy and Astrophysics, Australian National University, Canberra, ACT Australia
$^{32}$European Space Astronomy Centre, ESA, Villanueva de la Ca\~nada, Madrid, Spain\\
$^{33}$Dept. of Physics and Astronomy, University of Alabama, Tuscaloosa, AL\\
$^{34}$Dept. of Physics and Astronomy, State University of New York at Geneseo, Geneseo, NY\\
$^{35}$Observatoire de Geneve, University of Geneva, Geneva, Switzerland\\
$^{36}$Dept. of Astronomy, Columbia University, New York, NY\\
$^{37}$Dept. of Physics and Astronomy, The Johns Hopkins University, Baltimore, MD\\
$^{38}$Department of Physics and Astronomy, Bologna University, Bologna, Italy\\
$^{39}$Instituto de Astronomia, Universidad Nacional Autonoma de Mexico, Unidad Acad\'emica en Ensenada, Km 103 Carr. Tijuana-Ensenada, Ensenada 22860, Mexico \\
}

\maketitle

\begin{abstract}

We present the star cluster catalogs for 17 dwarf and irregular galaxies in the $HST$ Treasury Program ``Legacy ExtraGalactic UV Survey" (LEGUS). Cluster identification and photometry in this subsample are similar to that of the entire LEGUS sample, but special methods were developed to provide robust catalogs with accurate fluxes due to low cluster statistics. The colors and ages are largely consistent for two widely used aperture corrections, but a significant fraction of the clusters are more compact than the average training cluster. However, the ensemble luminosity, mass, and age distributions are consistent suggesting that the systematics between the two methods are less than the random errors. When compared with the clusters from previous dwarf galaxy samples, we find that the LEGUS catalogs are more complete and provide more accurate total fluxes. Combining all clusters into a composite dwarf galaxy, we find that the luminosity and mass functions can be described by a power law with the canonical index of $-2$ independent of age and global SFR binning. The age distribution declines as a power law, with an index of $\approx-0.80\pm0.15$, independent of cluster mass and global SFR binning. This decline of clusters is dominated by cluster disruption since the combined star formation histories and integrated-light SFRs are both approximately constant over the last few hundred Myr. Finally, we find little evidence for an upper-mass cutoff ($<2\sigma$) in the composite cluster mass function, and can rule out a truncation mass below $\approx10^{4.5}$M$_{\odot}$ but cannot rule out the existence of a truncation at higher masses.



\end{abstract}


\section{INTRODUCTION}
Dwarf galaxies are interesting laboratories with which to study the process of star formation. The extreme environments found in dwarfs (low-mass, low-metallicity, and low-star formation rate (SFR)) can provide leverage to test observational scaling relationships and predictions from theoretical models. Star clusters can be especially conspicuous in dwarf galaxies and a prominent tracer of the star formation process. For example, young massive clusters (M$>$10$^5~M_{\odot}$) are the products of extreme star formation periods \citep[i.e., high star formation efficiencies (SFE) $>$60\%;][]{turner15} and have been found in several local bursting dwarf galaxies \citep{billett02,johnson03,johnson04,calzetti15}. In addition, several cluster properties appear to scale with their host-galaxy
properties \citep[e.g., the brightest cluster and the total number of clusters, etc.;][]{larsen02,whitmore03,goddard10} that can provide clues to the physics of star formation.

Despite the important environmental conditions found in dwarf galaxies, their star cluster populations are often challenging to study due to the low SFRs and consequently the reduced numbers of clusters. The low number statistics can add scatter to established cluster-host relationships and cause difficulties in interpreting the results even in larger samples of dwarf galaxies \citep{cook12}. Thus, providing a large sample of dwarf galaxies whose star clusters have been uniformly identified and their properties uniformly derived can provide key insights into the star formation process.

There are two key factors that can act to reduce the amount of scatter found in the cluster-host relationships of dwarf galaxies are: 1) uniform identification of a more complete sample of clusters, and 2) measuring accurate total fluxes and consequently more accurate cluster ages and masses. There are other factors that can affect the accuracy of derived clusters properties (e.g., single stellar population model uncertainties, stochastic effects for low-mass clusters, and reddening law uncertainties); however, if implemented uniformly across a sample of clusters, then the scatter introduced by these other factors will be reduced.

The first factor that can add scatter into cluster-host relationships in dwarf galaxies is cluster identification. Traditionally, clusters have been identified by visually inspecting images to produce a cluster catalog. However, this method can result in missed clusters and contain biases depending on what an individual might identify as a cluster, which can depend on the size, shape, color, and luminosity of the cluster as well as the background and crowding environments nearby. Automated methods of cluster identification \citep{bastian12a,whitmore14,adamo17} could improve the completeness of clusters as these methods can flag all extended objects in the galaxy as cluster candidates. Unfortunately, these candidates require vetting by human classifiers where, in some cases, the number of candidates can be large compared to the number of real clusters. However, the number of candidates produced in dwarfs will likely be small, thus, making an automated identification method (with subsequent human vetting) practical in dwarf galaxies.

The second factor that can add cluster-host relationship scatter is inaccurate total fluxes and cluster properties. It is clear that high resolution imaging is required across at least 4 photometric bands to obtain clean photometry and accurate physical properties of clusters in nearby galaxies \citep{anders04a,bastian14}. Even with high resolution imaging, the aperture correction used for clusters can produce significantly different total fluxes \citep{chandar10b} and the subsequently derived physical properties (age, mass, and extinction).

In this paper we examine various methods to identify and measure total fluxes of clusters in a sample of 17 dwarf and irregular galaxies from the Legacy ExtraGalactic UV Survey \citep[LEGUS;][]{legus} where high-resolution HST images in 5 bands have been acquired by LEGUS for each of these dwarfs. We focus on the comparisons between automated and human-based identification as well as which photometric methods produce accurate total cluster fluxes and consequently produce accurate physical properties (i.e., age, mass, and extinction). We then compare our results of identification and photometry with those from previous cluster studies in dwarf galaxies. Finally, we conclude by examining the basic properties of clusters in these extreme environments and test if these properties change with galaxy-wide properties.


\input{tables/GlobalProp.tex}

\section{Data \& Sample\label{sec:data}}
In this section we describe the LEGUS dwarf and irregular galaxy sub-sample and how it compares to the full LEGUS sample. The data and properties of the entire LEGUS sample are fully described in \citet{legus}, but we provide an overview here.  The LEGUS sample consists of 50 nearby galaxies within a distance of 12~Mpc to facilitate the study of both individual stars and star clusters. A combination of new WFC3 and existing ACS HST imaging constitute the LEGUS data resulting in 5 bands for each galaxy that cover near UV and optical wavelengths. The HST filters available in the LEGUS galaxies are F275W, F336W, F438W/F435W, F555W/F606W, and F814W, and are hereafter referred to as $NUV$, $U$, $B$, $V$, $I$, respectively. The global properties of the full LEGUS sample span a range in SFR ($-2.30<\rm{log(SFR; M_{\odot}yr^{-1})}<0.84$), stellar mass ($7.3<\rm{log(M_{\star}; M_{\odot})}<11.1$), and SFR density ($-3.1<\rm{log(\Sigma_{\rm SFR}; M_{\odot}yr^{-1}kpc^{-2})}<-1.5$). The normalized areas used for SFR densities are the D25 isophotal ellipses from RC3 \citep{rc3} as tabulated by NED\footnote{https://ned.ipac.caltech.edu}.

\begin{figure*}
  \begin{center}
  \includegraphics[scale=0.7]{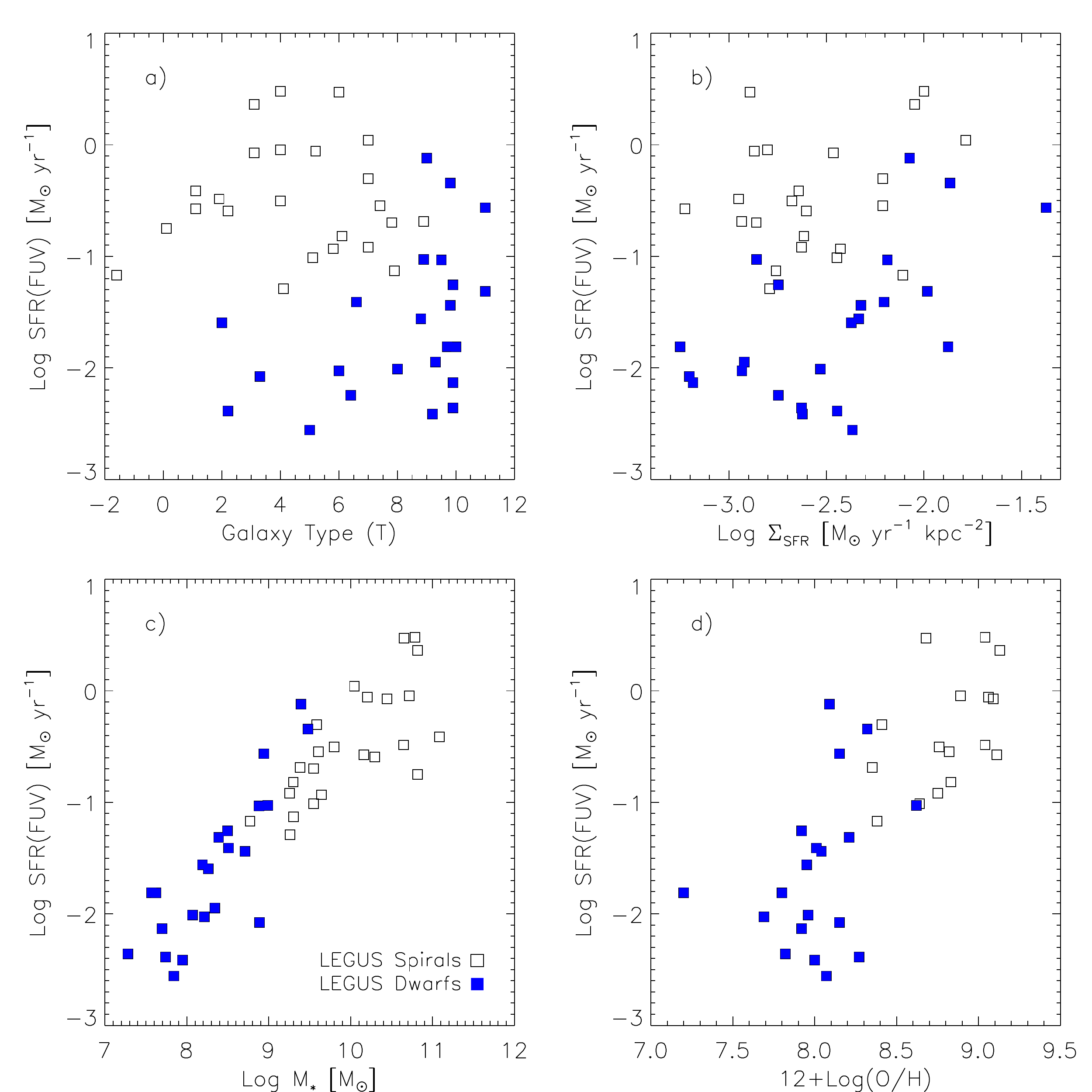}
  \caption{A four-panel plot showing a comparison of physical properties between the LEGUS dwarfs and spirals. Each plot shows the global SFR versus morphology, SFR density, stellar mass, and gas-phase metallicity. The LEGUS dwarfs tend to have lower SFRs, SFR densities, stellar masses, and metallicities.}
   \label{fig:genprop}
   \end{center}
\end{figure*}

The dwarf and irregular galaxy sub-sample was chosen based on the absence of obvious spiral arms and dust lanes in the HST color images. As a result of this morphological selection, the galaxies studied here have irregular morphologies and may not strictly be considered dwarf galaxies. However, the majority (15 out of 17) have stellar masses below log($M_{\star}\leq9~\rm{M}_{\odot}$) (see Figure~\ref{fig:genprop}). There are 23 galaxies in the LEGUS sample that meet the morphological criteria, and all 23 global galaxy properties are presented in the subsequent paragraphs of this section. However, in the cluster analysis sections of this study (\S\ref{sec:clustcat} and beyond) we utilize only the 17 dwarfs available in the public cluster catalog release of June 2018. 

We used published physical properties to verify that the dwarf sample ($N$=23) tended to have low SFRs, low stellar masses (M$_{\star}$), and low metallicities; these properties are presented in Table~\ref{tab:genprop}. The FUV-derived SFRs are taken from \citet{lee09b} and \citet{legus}. The stellar masses are taken from \citet{cook14c} and are computed from mass-to-light ratios of the Spitzer 3.6$\mu m$ fluxes from \citet{dale09}. The metallicities are taken from the compilation of \citet{cook14c}, where oxygen abundances derived from direct methods are favored, but do contain strong-line measurements when direct method values were not available. The global properties of the dwarf sub-sample span a range in SFR ($-2.30<\rm{log(SFR; M_{\odot}yr^{-1})}<-0.03$), stellar mass ($7.3<\rm{log(M_{\star}; M_{\odot})}<9.5$), and SFR density ($-3.1<\rm{log(\Sigma_{\rm{SFR}}; M_{\odot}yr^{-1}kpc^{-2})}<-1.5$); note that the dwarfs span the full range of $\Sigma_{\rm{SFR}}$ as the full LEGUS sample.

Figure~\ref{fig:genprop} illustrates the galaxy-wide physical properties of both the full LEGUS sample and the dwarf sub-sample. Panel `a' shows the distribution of the SFR versus galaxy morphological type ($T$), where the dwarf sub-sample tends to populate the later-type and lower SFR range of the entire sample. Panel `b' shows the distribution of the SFR versus $\Sigma_{\rm{SFR}}$, where the dwarf sub-sample spans a large range of $\Sigma_{\rm{SFR}}$. Panels `c' and `d' show the distribution of SFR versus stellar mass and metallicity, respectively. The dwarf sub-sample tends to have lower stellar mass and metallicity.

\section{Cluster Catalogs}\label{sec:clustcat}
In this section we describe how the cluster catalogs for the LEGUS dwarf galaxies are constructed. The procedures used to produce the cluster catalogs follow that of \citet{adamo17}, and involve multiple steps: detection of candidates, classification, photometry and extrapolation to total flux, and SED fitting to obtain physical properties (e.g., age, mass, and extinction). 

One of the main goals of the LEGUS project is to determine whether the properties of star clusters are dependent on the galactic environment where they live. Dwarf galaxies offer a environment distinct from more massive spiral galaxies in which to explore this possibility.  Given the possibility that star clusters in dwarfs may exhibit different properties, it is not unreasonable to assume that care must be taken when applying the methods of cluster detection and characterization developed using the LEGUS spiral galaxies to dwarfs galaxies to ensure that systematics are not introduced.  In this section we highlight two additional steps that were taken to check for possible systematics: a visual search for clusters, and an alternate method to compute total cluster fluxes.





\subsection{CLUSTER IDENTIFICATION \label{sec:clustid}}

\subsubsection{Automated Cluster Candidate Detection \label{sec:autoID}}
The LEGUS cluster pipeline allows the user to tailor parameters for selection and photometry to appropriate values for each galaxy. The pipeline begins by utilizing SExtractor \citep{sex96} to identify point and point-like objects in the V-band image to create an initial catalog of both stars and star cluster candidates. 

A key step in the overall process is the visual identification of isolated stars and star clusters which serve as training sets to guide separation of clusters from stars, and to determine appropriate photometric parameters (e.g., aperture radius and aperture correction; see \S~\ref{sec:phot}).  Stars and star clusters are separated based on the extent of their radial profiles as measured by the concentration index (CI). In the LEGUS cluster pipeline, CI is defined as the difference in magnitudes as measured from two radii (1 pixel minus 3 pixels). For each galaxy, a CI separation value is chosen by the user via comparison of CI histograms for training stars and clusters, where the high end of the stellar CI histogram helps to set the stellar-cluster CI threshold. Typical CI thresholds in the LEGUS dwarfs are 1.2 to 1.4~mag, similar to the values used for the LEGUS spirals.



After cluster candidates are identified, a final cut is made by the LEGUS pipeline after the photometry is completed (see \S\ref{sec:phot}) where sources with an absolute magnitude fainter than $M_V{=}-6$ are excluded. Previous studies have shown that separation of stars and clusters in absolute magnitude occurs in the range of $M_V{=}-6$ to $M_V{=}-8$~mag, where stars can be as bright as $M_V{=}-8$~mag and clusters can be identified as faint as $M_V{=}-6$~mag \citep{larsen04,chandar10b}. Thus, a $M_V{=}-6$~mag cut is employed by the LEGUS pipeline to remove potential stellar contamination while minimizing the loss of potential star clusters. 

\input{tables/ClustProp.tex}

The total number of cluster candidates found in the LEGUS dwarfs with this process is 3475. The number of candidates per galaxy is presented in Table~\ref{tab:clustprop} and spans from over a thousand in NGC4449 to 18 in NGC5238.


\subsubsection{Classification of Cluster Candidates \label{sec:visclass}}
After the automated detection is complete, the resulting sources are vetted for contaminants (background galaxies, stars, artifacts, etc) and classified based on morphology and symmetry.

\begin{figure}
  \begin{center}
  \includegraphics[scale=1.35]{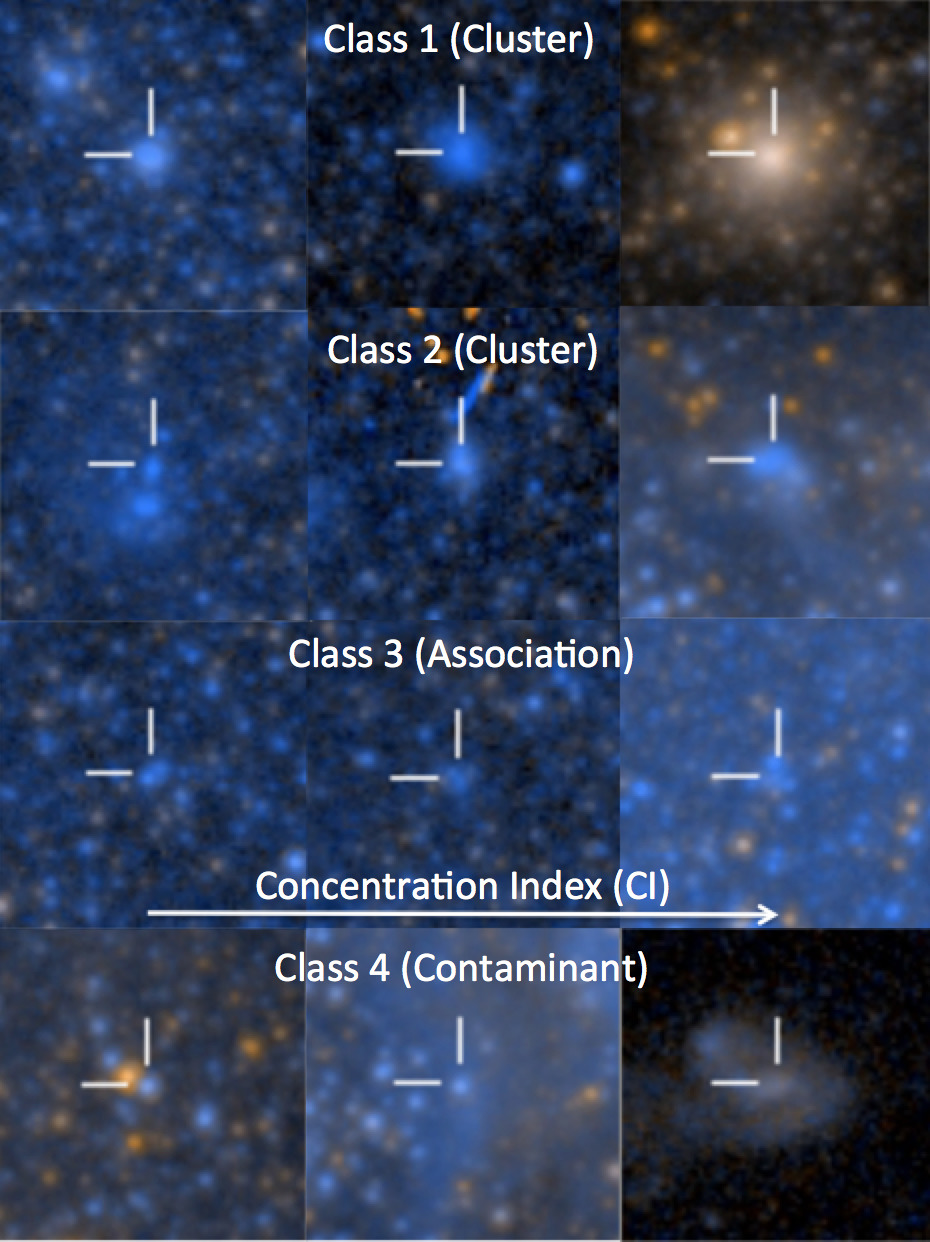}
  \caption{An HST color mosaic of example clusters. The 4 rows present 3 examples for class 1,2,3,4, from top-to-bottom,  where the 3 examples across the row represent sources with low CI values (compact), average, and large CI values. The class 4 examples across the bottom row represent stars with nearby contamination, a star with contaminating nebular emission, and a background galaxy.}
   \label{fig:manidmosaic}
   \end{center}
\end{figure}  

A full description of the classification scheme can be found in \citet{adamo17} and Kim et al. (2018; in prep), but we provide a brief overview here. Classifications were performed by at least 3 LEGUS team members, where the final classification is defined as the mode of all classifications. Class 1 sources are those that have extended radial profiles with spherical symmetry. Class 2 sources are those that are extended, but have some degree of asymmetry in their radial profiles. Class 3 sources are those with multiple peaks in their radial profiles. Class 4 sources are those considered to be contaminants (e.g., obvious stars, background galaxies, random overdensities of nebular emission, etc.). The morphologies potentially provide insight into the evolutionary status of the clusters.  Class 1 and 2 sources may be gravitationally bound star clusters while class 3 sources (showing multiple stellar peaks) are referred to as compact stellar associations, which may be in the process of being disrupted \citep{grasha15,adamo17}. 

Figure~\ref{fig:manidmosaic} shows the HST color image cutouts of three example clusters for classes 1, 2, 3, and 4 from the top to the bottom panels. Examples from left-to-right in Figure~\ref{fig:manidmosaic} show representative CI values near the minimum, average, and maximum for class 1, 2, and 3. The class 4 examples in Figure~\ref{fig:manidmosaic} from left-to-right show a star with a nearby contaminating object, a star that is spatially coincident with an overdense nebulous region, and a background galaxy, respectively. The majority of class 4 cluster candidates are stars whose CI values are inflated due to light from nearby sources.

Integrated over the LEGUS dwarf galaxy sample, the cluster pipeline finds 944, 495, and 2036 sources for classes 1-2, 3, and 4, respectively (i.e., the majority are determined to be contaminants). The number of confirmed candidates in each class in individual galaxies is presented in Table~\ref{tab:clustprop}.


\subsubsection{Visual Cluster Search \label{sec:manID}}

A visual search of the HST color images was also performed to provide a check on the LEGUS cluster pipeline. One of the authors (DOC) used images created from the $V$- and $I$-bands to search for clusters in the LEGUS dwarfs using procedures similar to \citet{cook12}.  Clusters were identified as: a close grouping of stars within a few pixels with an unresolved component, or a single extended source with spherical symmetry. Sources exhibiting evidence of spiral structure (indicating a background galaxy) were excluded.  The clusters were subsequently classified by multiple LEGUS team members as class 1, 2, or 3. 

\begin{figure}
  \begin{center}
  \includegraphics[scale=0.48]{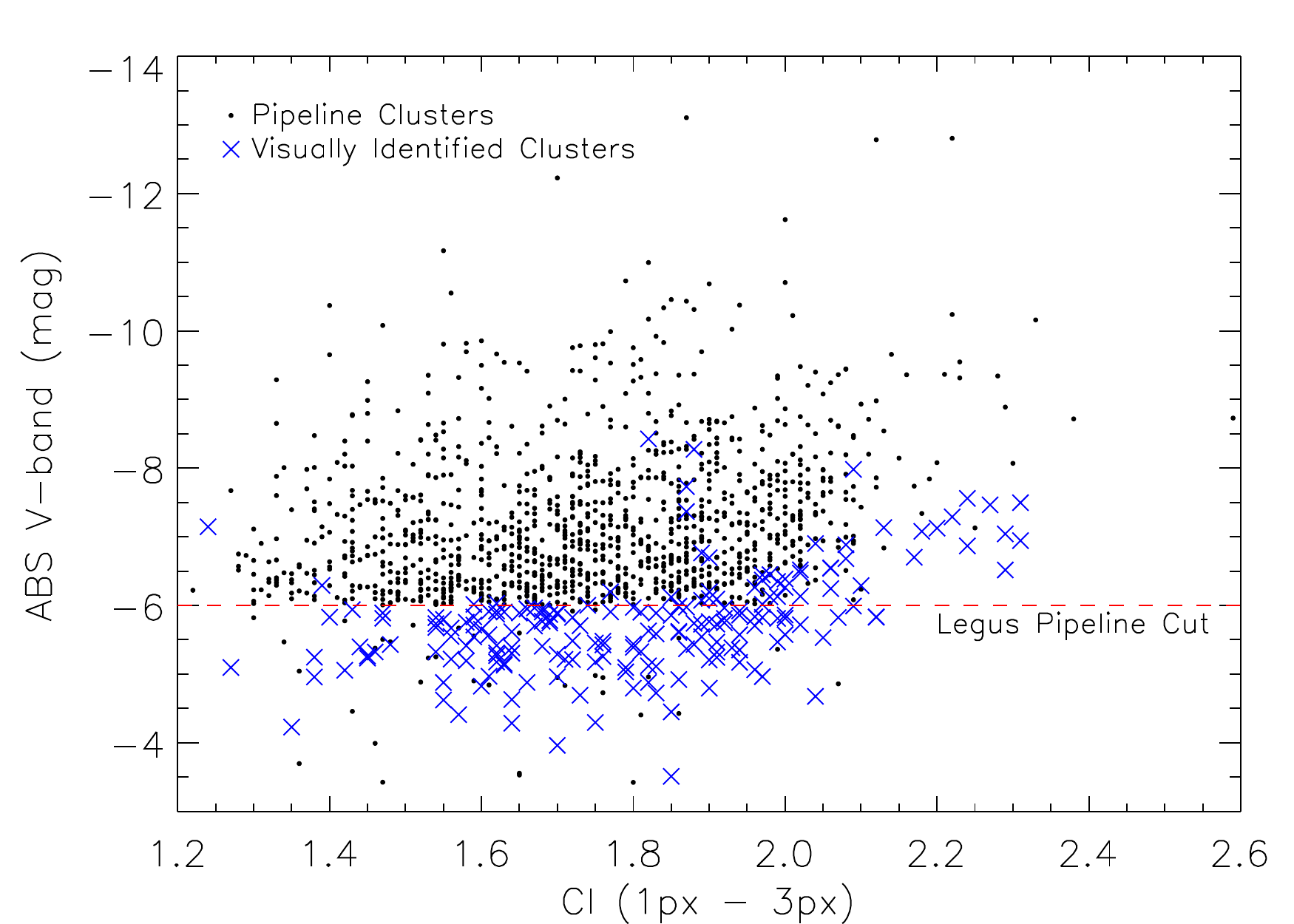}
  \caption{The absolute $V-$band magnitude of the clusters found in the LEGUS dwarf galaxy sub-sample. The blue X's represent the human-based clusters not found by the LEGUS cluster pipeline. The majority of these clusters are fainter than the $M_V{=}-6$~magnitude cut imposed by the pipeline.}
   \label{fig:absci}
   \end{center}
\end{figure}

In total, 193 clusters were found in the visual search that were missing from the catalog produced by the LEGUS extraction tool, and these were added to the LEGUS catalogs.\footnote{For users of the LEGUS cluster catalogs, the clusters missed by the LEGUS pipeline are indicated with a 'manflag' value equal to one.}  Figure~\ref{fig:absci} is a plot of absolute $V-$band magnitude versus CI for clusters found via the LEGUS pipeline and visual inspection clusters missed by the pipeline. We find that the majority (74\%) of clusters missed by the LEGUS pipeline are fainter than the $M_V=-6$ cut imposed by the pipeline. In addition, we find that the LEGUS cluster pipeline successfully recovers the majority (88\%) of clusters to its stated limits (i.e., those brighter than $M_V=-6$~mag). 

The small number of visually identified clusters brighter than $M_V=-6$ that were missed by the LEGUS pipeline can be explained by: user defined limits, the 3 sigma detection limit imposed by the pipeline, or poor source extraction in high density environments. The compact cluster missed at CI$\sim$1.25, $M_V=-7.2$~mag was cut in the pipeline due to the user imposed CI cut (=1.3). The missing clusters with the highest CI values (CI$>$2.1) were missed due to larger photometric errors just above the LEGUS pipeline detection threshold of 0.3~mag (i.e., low surface brightness). The remaining handful of clusters are located in a rapidly varying background region in NGC4449. All of these missing clusters were added into the final catalog.




\subsection{Cluster Photometry \label{sec:phot}}
In this section we describe the procedure to utilize training clusters to determine the radius at which we perform aperture photometry on the clusters, and the aperture correction to obtain total fluxes. These two photometry parameters can greatly affect the total flux of each cluster, and consequently the derived physical properties. 

It is relatively straightforward to determine the aperture radius, which is based on the normalized radial flux curves of the training clusters, where the median radial profile provides the profile of a typical star cluster in each galaxy. The aperture is chosen to be the radius at which 50\% of the median flux is contained within the aperture. The aperture radii values allowed in our analysis here have discrete values of 4, 5, and 6 pixels. 

The challenge in this process particular to dwarf galaxies is that galaxies with low SFRs have small populations of clusters overall, and the isolated clusters that can be used for the training set may be few.  Table~\ref{tab:apcorr} shows the number of training clusters found in each of the dwarf galaxies, where the number ranges from 2 to 55.  It is possible that this could lead to aperture corrections that are not well determined for low SFR galaxies, so we have investigated two methods: 1) an average aperture correction as measured from the isolated clusters in the training set for each galaxy, and 2) a correction based the measured CI of each cluster, where the correction to total flux is derived from a suite of artificial star clusters embedded in our HST imaging.

\subsubsection{Average Aperture Correction}
\begin{figure*}
  \begin{center}
  \includegraphics[scale=0.5]{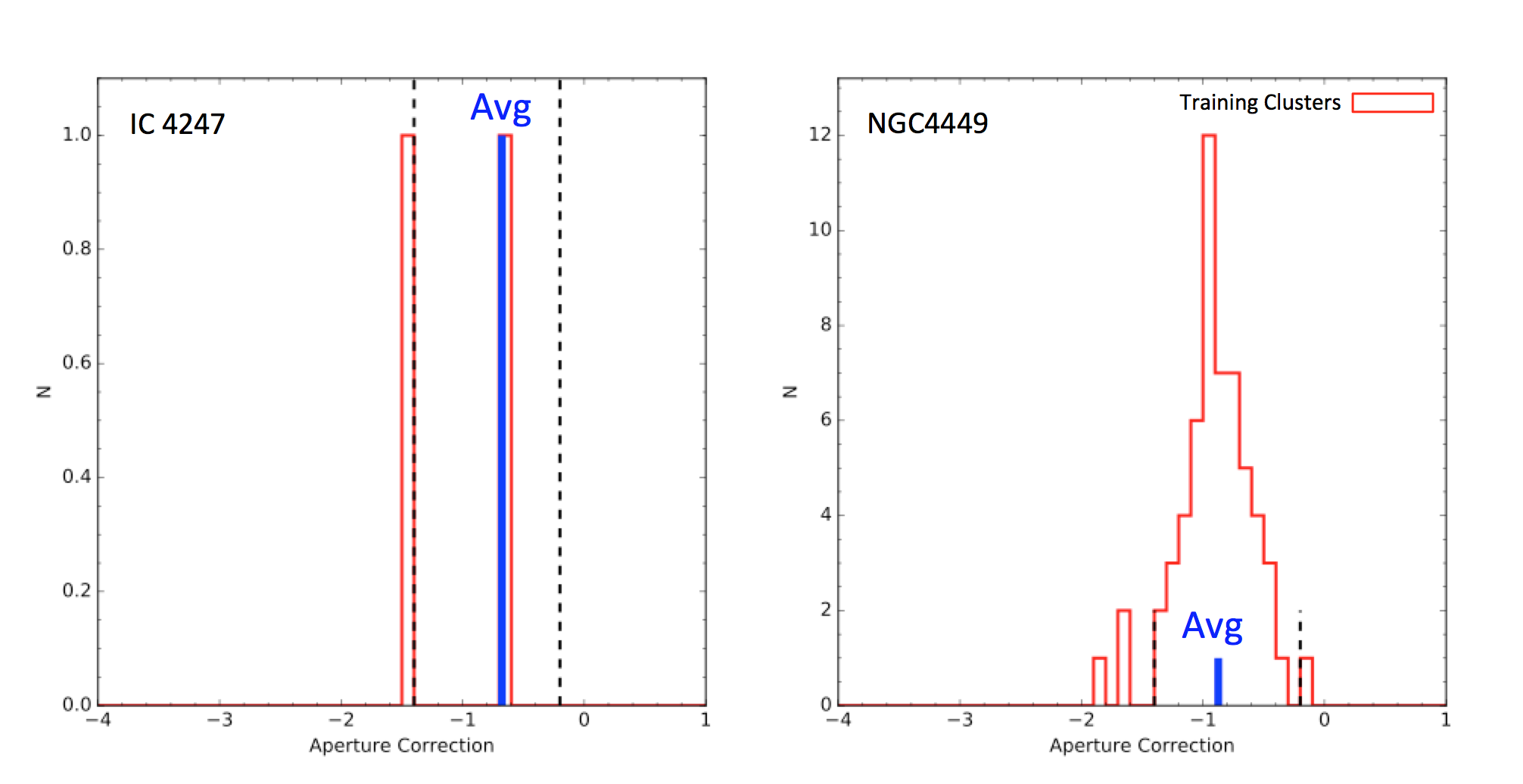}
  \caption{The aperture correction histograms measured for each training cluster in the $V-$band for IC4247 (left) and NGC4449 (right). The vertical dashed lines represent limits imposed on the average aperture correction to exclude outlier training clusters. With only 1 training cluster in IC4247 (since the other is outside the allowed region), the average aperture correction relies on a single measurement resulting in an uncertain average aperture. The numerous training clusters in NGC4449 provide a well-behaved aperture correction histogram resulting in a robust average aperture correction. }
   \label{fig:apcorrcomp}
   \end{center}
\end{figure*}

The first method adopts an average aperture correction of training clusters. This method has been widely used \citep{chandar10b,adamo17} and has the advantage of being resistant to outliers in the training set. Here, we take the difference in magnitudes measured in a 20 pixel radius aperture minus that measured in the ``half light" aperture (i.e., a 4,5,6 pixel radius as determined above) as the correction. Figure~\ref{fig:apcorrcomp} presents the aperture correction histogram in the V-band for the training clusters in IC4247 (panel a) and NGC4449 (panel b), which shows that the average aperture correction is not well defined for galaxies with low numbers of training clusters.


The average aperture correction in NGC4449 follows the peak of the histogram, thus recovering the aperture correction of a typical cluster. However, the histogram in IC4247 is not well determined as there are only 2 training clusters with corrections that differ by a factor of $\sim$2 ($\sim$0.75~mag). Furthermore, since 1 of the 2 training clusters in IC4247 lies outside the allowed limits of the aperture correction histogram, the average aperture correction is based on only a single training cluster making the aperture correction highly uncertain. Depending on the radial profile of only 1 or 2 training clusters the average aperture correction may not reflect a typical cluster in a dwarf galaxy, and may in fact cause this correction to vary across filters within a galaxy.


\input{tables/ApCorr.tex}

\begin{figure}
  \begin{center}
  \includegraphics[scale=0.53]{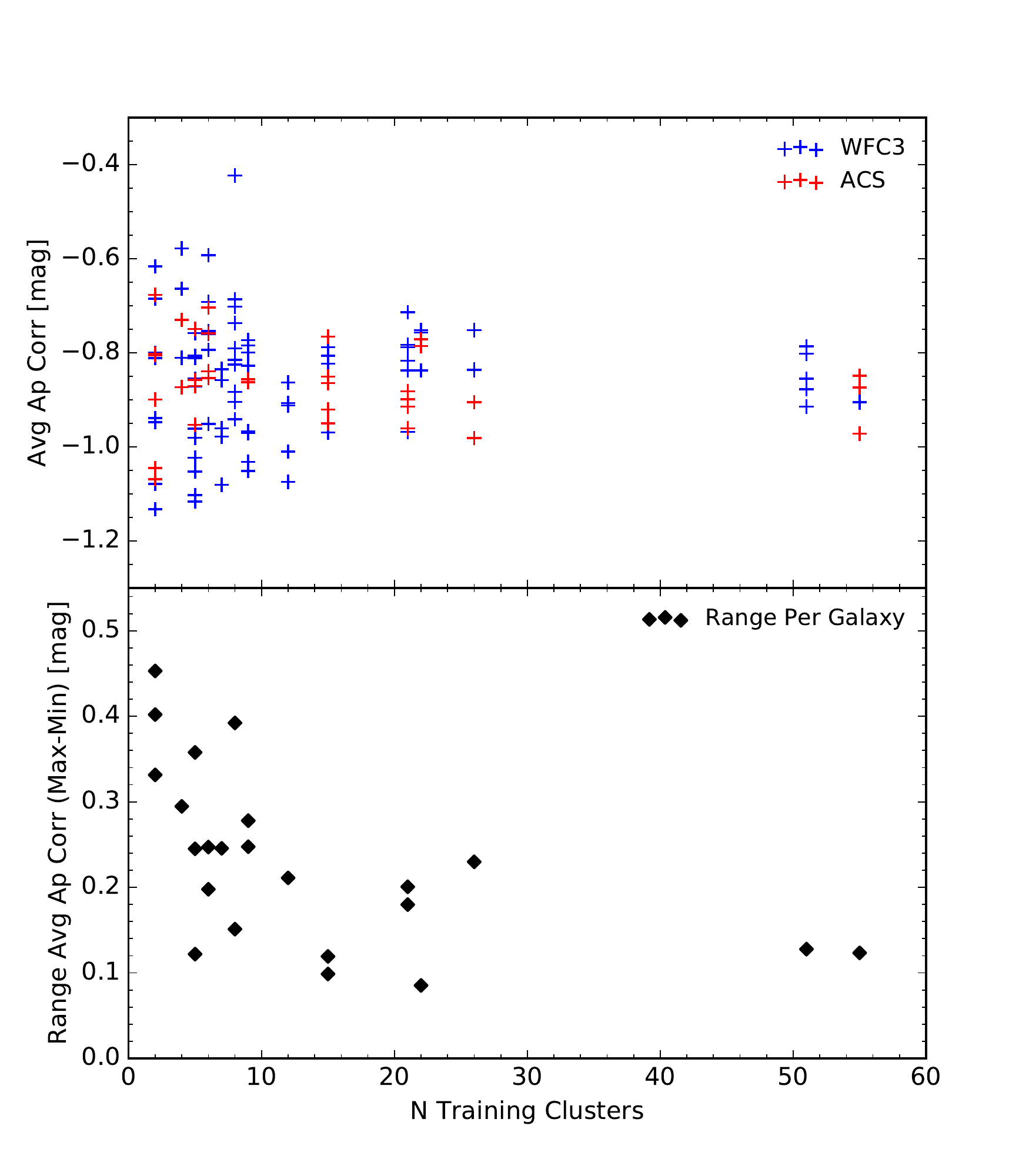}
  \caption{Top panel: the average aperture correction for all filters in the LEGUS dwarf galaxies plotted against the number of training clusters used to derived the average aperture correction. Some of the y-axis shifts can be accounted for by different photometry radii. However, there exists large aperture correction spreads (0.4~mag) for individual galaxies with fewer numbers of training clusters. Bottom panel: the range in average aperture correction across the filters in each galaxy. The range increases significantly below 10 training clusters.}
   \label{fig:apcorrN}
   \end{center}
\end{figure}

For example, Table~\ref{tab:apcorr} shows the average aperture corrections in each filter across the LEGUS dwarf galaxies, where column 10 gives the range in the aperture correction across the filters (Columns 2$-$9). The range in aperture corrections for a single galaxy is as low as 0.09~mag and as high as 0.45~mag. The top panel of Figure~\ref{fig:apcorrN} graphically presents the average aperture corrections for all filters in each galaxy versus the number of training clusters. The bottom panel of Figure~\ref{fig:apcorrN} demonstrates that there is a larger spread of aperture corrections in galaxies with lower numbers of training clusters (N$<$10). We note that we found no correlations between distance and the average aperture corrections, the range in aperture corrections, nor the number of training clusters.


The larger scatter in the average aperture corrections at lower numbers of training clusters may artificially change the shape of a cluster's SED making the cluster more blue/red which can affect the derived age and extinction. In the next section, we explore a second aperture correction based on the CI of each cluster to mitigate the effects of low training cluster numbers on the aperture corrections.

\subsubsection{CI-based Aperture Correction \label{sec:ciapcorr}}
An alternative method used to derive an aperture correction is based on the radial profile of each cluster as quantified by the concentration index (CI) in each filter. This method has also been widely used in the literature \citep{chandar10b,bastian12a,adamo15}, but can have drawbacks where uncertain aperture corrections can be found for faint sources with marginal detections in some filters.


We derive a relationship between the aperture correction and the CI for model clusters in each filter image. The model clusters are generated using the MKSYNTH task in the BAOLAB package \citep{baolab} following the procedure of \citet{chandar10b} where different sized clusters are constructed by convolving a KING30 profile \citep{king66} of various FWHM values with an empirically-derived stellar PSF made from isolated stars found in each image \citep[see also;][]{anders06}. The PSF sizes for the WFC3 and ACS cameras are 2.1 pixels (0.105\arcsec) and 2.5 pixels (0.125\arcsec), respectively.  Model clusters are then injected into relatively sparse regions of all filter images for several LEGUS galaxies (both dwarfs and spirals). In addition, we inject the empirically-derived PSF into these same regions to define the expected CI threshold  between stars and star clusters. After injecting both model clusters and model stars (i.e., the empirical PSF) into each image, we extract the resulting photometry using the ``half light" apertures and measure the CI and aperture correction. We note that King profiles are often used for globular clusters (i.e., self-gravitating systems) and that younger clusters show better empirical fits to Moffat profiles \citep{elson87}. However, we find no difference in the aperture correction--CI relationships for both King and Moffat profiles at low and high CI values and little difference (0.1--0.2~mag) at intermediate CI values (1.5--1.7).

\begin{figure}
  \begin{center}
  \includegraphics[scale=0.49]{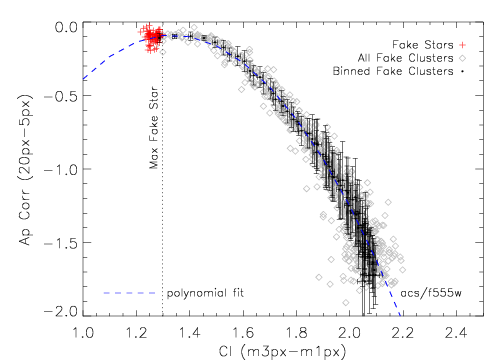}
  \caption{The aperture correction measured for model stars (red pluses) and clusters (grey diamonds) plotted against the measured CI, where the curved dashed line represents the polynomial fit to both model stars and clusters.The black filled circles represent the median and standard deviation of all model clusters in CI bins. The dashed-blue line in Figure~\ref{fig:ciapcorrex} represents the cubic polynomial fit to both the model stars and the median extracted model clusters.}
   \label{fig:ciapcorrex}
   \end{center}
\end{figure}

Figure~\ref{fig:ciapcorrex} shows a plot of the aperture correction versus CI for model clusters and stars inserted into one of the ACS-F555W filter images, where we find the expected relationship in that higher CI values (i.e., more extended) have larger aperture corrections (i.e., more negative). The cubic polynomial fit to both the model stars and clusters is consistent with previous studies \citep{chandar10b}. In addition, we find a model star-cluster CI threshold of 1.2 and 1.3~mag for the WFC3 and ACS cameras, respectively.

\begin{figure}
  \begin{center}
  \includegraphics[scale=0.115]{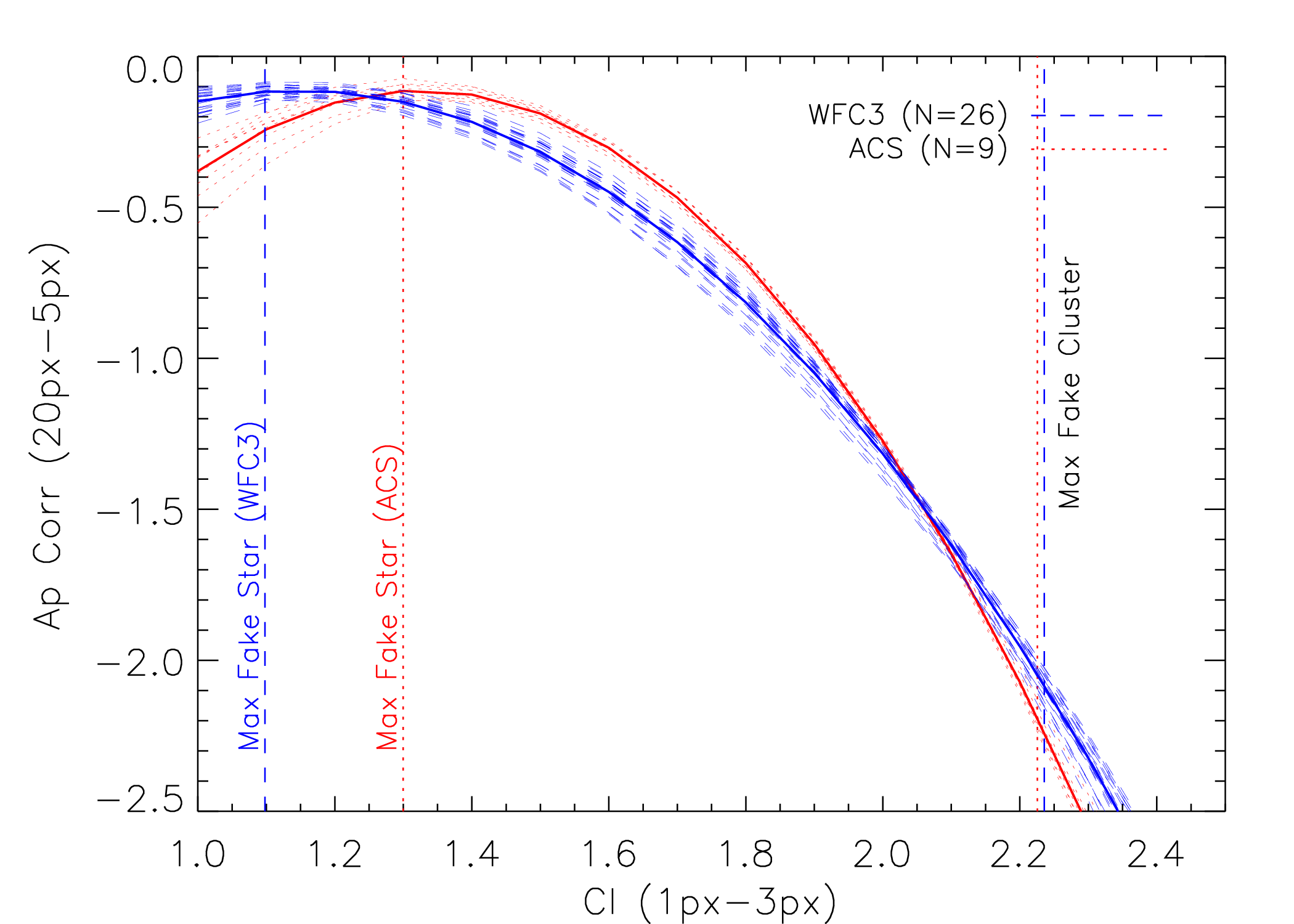}
  \caption{The aperture correction-CI polynomial fits for all filter-camera combinations for the 7 LEGUS galaxies used to derive these fits. The fits for each camera show good agreement across filters. Thus, the final CI-based aperture correction polynomial fits are given as the median of the polynomial fits in each camera (see Table~\ref{tab:fakeclust}). The vertical dotted and dashed lines represent the maximum measured model star CI (CI min) and the maximum measured model cluster CI (CI max) for the ACS and WFC3 cameras, respectively. These limits represent the range of CI values measured for model clusters.}
   \label{fig:ciapcorrall}
   \end{center}
\end{figure}

In total, we have injected model stars and clusters into the images of 7 LEGUS galaxies (4 spirals and 3 dwarfs) whose imaging contains all camera-filter image combinations present in the entire LEGUS survey. We derive polynomial relationships between aperture correction and CI for all camera-filter image combinations since the WFC3 and ACS PSFs are different. Figure~\ref{fig:ciapcorrall} shows the polynomial fits for all images (WFC3 and ACS) in the 7 galaxies where we find similar polynomial fits for each camera (regardless of the filter). Thus, we have defined a single polynomial fit for each camera as the median of the fits for all filters, which are represented as solid lines in Figure~\ref{fig:ciapcorrall}.

Finally, we repeat the analysis for different aperture radii since the aperture correction will depend on the aperture radius. We derive a median relationship for each camera with the three aperture radii allowed in the LEGUS cluster pipeline (4, 5, 6 pixels). We do not show the aperture correction versus CI plots for the other 2 aperture radii since they are similar to that in Figure~\ref{fig:ciapcorrall}, but with shifted aperture corrections (i.e., y-axis). We present the cubic polynomial fits for all three apertures in Table~\ref{tab:fakeclust}

\input{tables/fakeclust.tex}

\begin{figure}
  \begin{center}
  \includegraphics[scale=0.53]{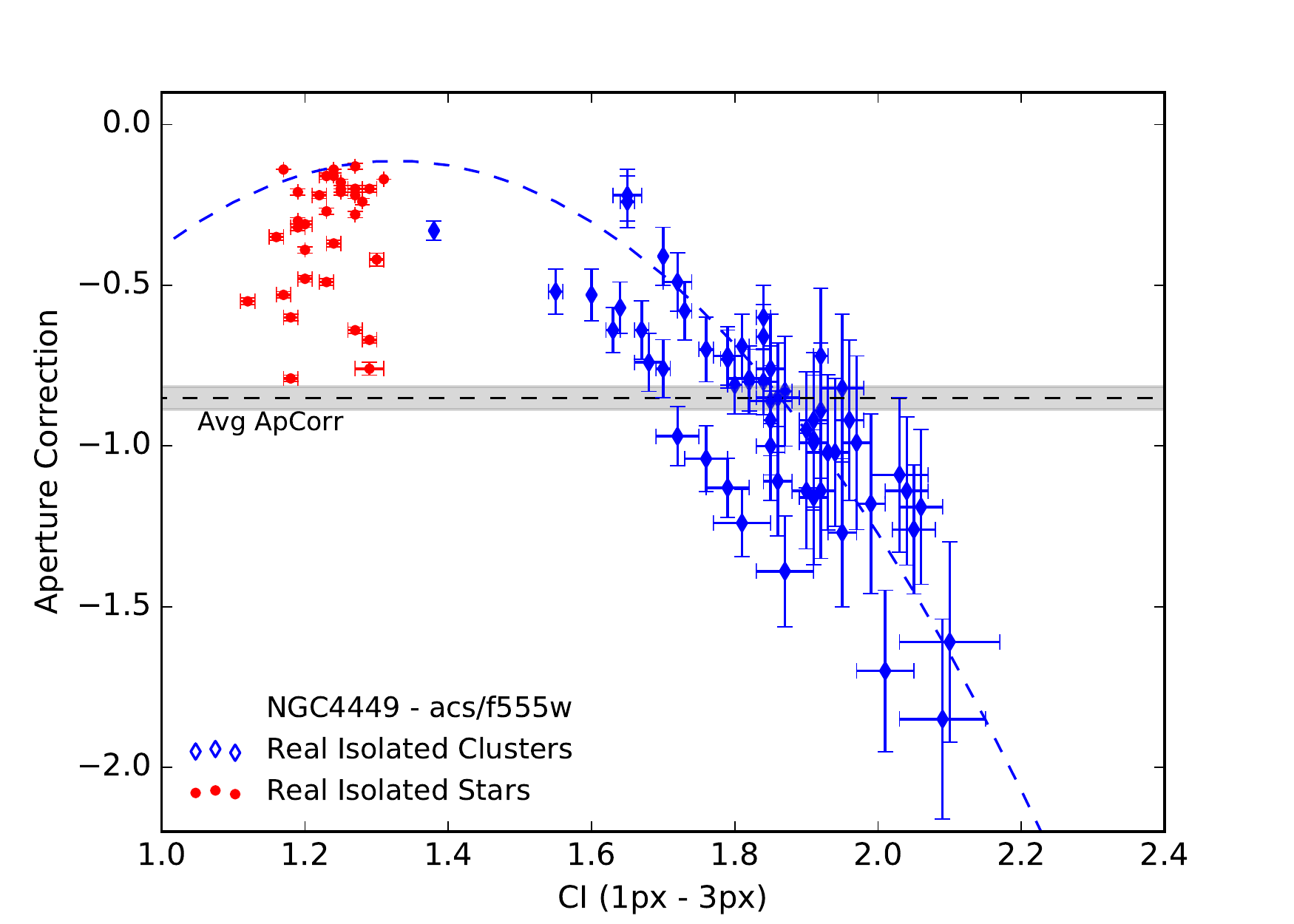}
  \caption{The measured aperture correction versus CI values for the training clusters in NGC4449. Both the training stars and clusters show good agreement with the polynomial fits derived from model stars and clusters. The horizontal dashed line represents the average aperture correction where the error is represented by the gray shaded area. The difference in average and measured aperture correction at low and high CI values can differ by as much as 1~magnitude. }
   \label{fig:ciapcorrreal}
   \end{center}
\end{figure}

As a check on the model cluster polynomial fits, we compare these fits to the aperture corrections and CI values for the real isolated stars and clusters of NGC4449 for the $V-$band in Figure~\ref{fig:ciapcorrreal}. We do not show the other filters since they show similar agreement with similar scatter. We find that the measured values of both the real stars and clusters show good agreement with the polynomial fit generated from model stars and clusters; including those clusters with large CI values near CI=2.1~mag. One of these clusters has an extended radial profile (CI=2.14~mag) and a measured CI-based aperture correction of --1.55~mag. The average aperture correction in this filter is --0.85~mag, which is a 0.7~mag difference. 

\begin{figure}
  \begin{center}
  \includegraphics[scale=0.45]{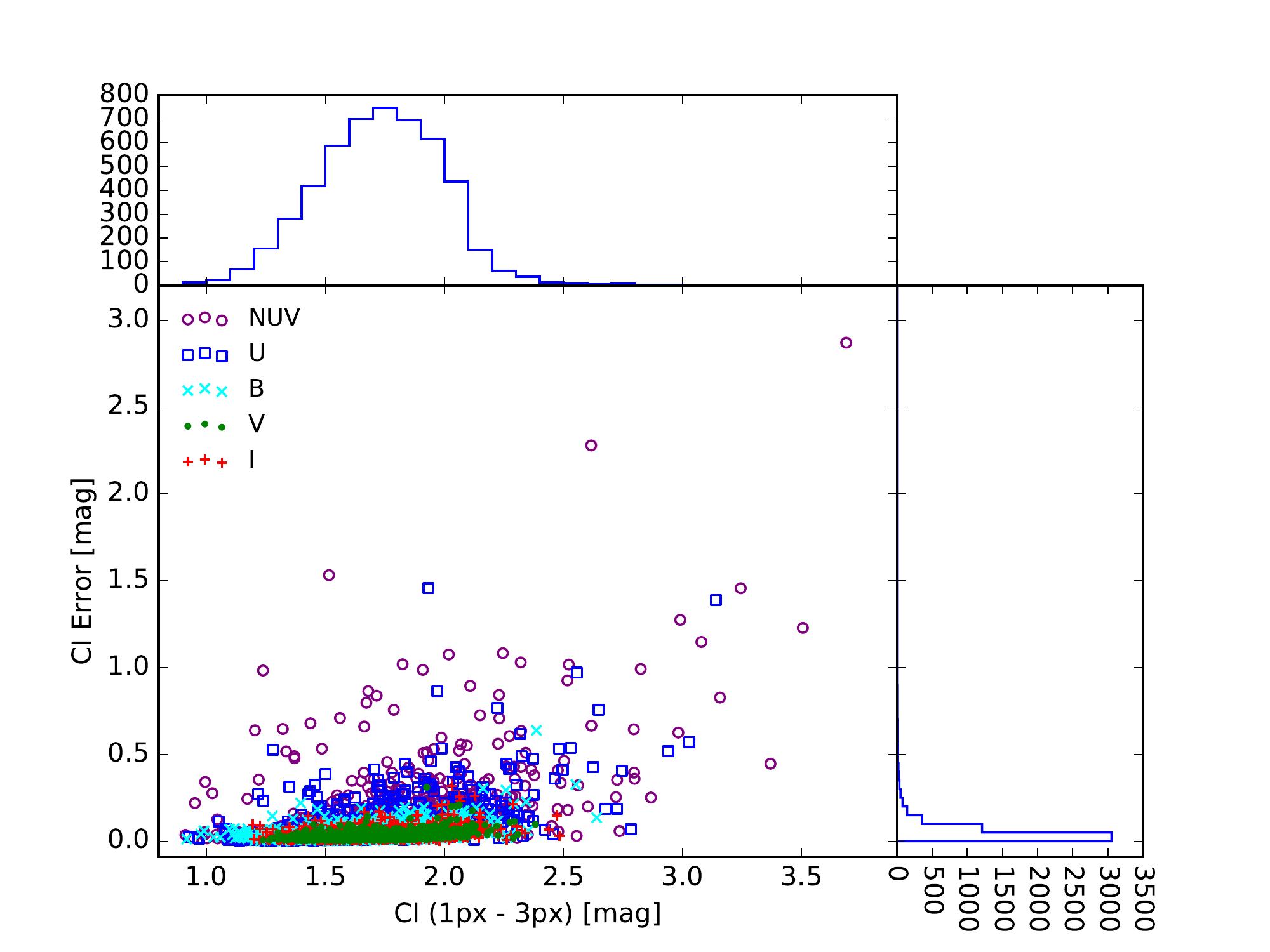}
  \caption{The measured CI and errors for all clusters in all filters in the LEGUS dwarfs. We find a median CI of 1.7~mag with a standard deviation of 0.27~mag, and a median CI error of 0.04~mag with a standard deviation of 0.12~mag. }
   \label{fig:CIerrCI}
   \end{center}
\end{figure}

The main panel of Figure~\ref{fig:CIerrCI} shows the CI values versus their uncertainties for all clusters in all filters, where each symbol represents the photometry information from the five filters. In addition, we plot the histograms for CI and CI error values in the top and right histogram panels, respectively. We find that the measured CI values of all clusters span a range in values where the median value is 1.7~mag with a standard deviation of 0.27~mag. We also find that the majority of the CI errors are relatively low, where the median value is 0.04~mag with a standard deviation of 0.12~mag. The low CI errors suggest that the majority of our CI-based aperture corrections are well defined and suitable for aperture corrections.


A caveat to using the CI-based aperture correction method is that large aperture correction uncertainties can exist for clusters with marginal detections in various filter images. We find that the CI errors increase sharply for the faintest clusters near 24~mag. Typically this occurs in our bluest filters ($NUV$- or $U$-band) due to the lower sensitivity of these observations and/or clusters with a redder SED. The fraction of clusters with a CI error greater than 0.1~mag is 34\% and 4\% for the $NUV-$ and $V-$band filters, respectively. Thus, the CI-based aperture corrections may change the SED shape of clusters with poor detections in some bands; this is more likely to occur for older/redder clusters.




\section{RESULTS}\label{sec:results}
In this section we provide a detailed comparison of the two aperture corrections and their effects on cluster colors and derived physical properties (age, mass and extinction). We also compare the LEGUS dwarf galaxy cluster catalogs to those previously identified in another large sample of dwarf galaxies \citep{cook12}. Finally, we present the luminosity, mass, and age distributions along with an investigation of observable cluster properties across galaxy environment. 

\subsection{CI-based versus Average Aperture Correction}


The goal of this section is to test what aperture corrections provide the most accurate total fluxes, colors, and physical properties (age, mass, and extinction) for clusters in dwarf galaxies. Here, we provide methodology guidelines for future cluster investigations in galaxies with small cluster populations.

\subsubsection{Photometric Property Comparison} \label{sec:photcomp}

\begin{figure}
  \begin{center}
  \includegraphics[scale=0.5]{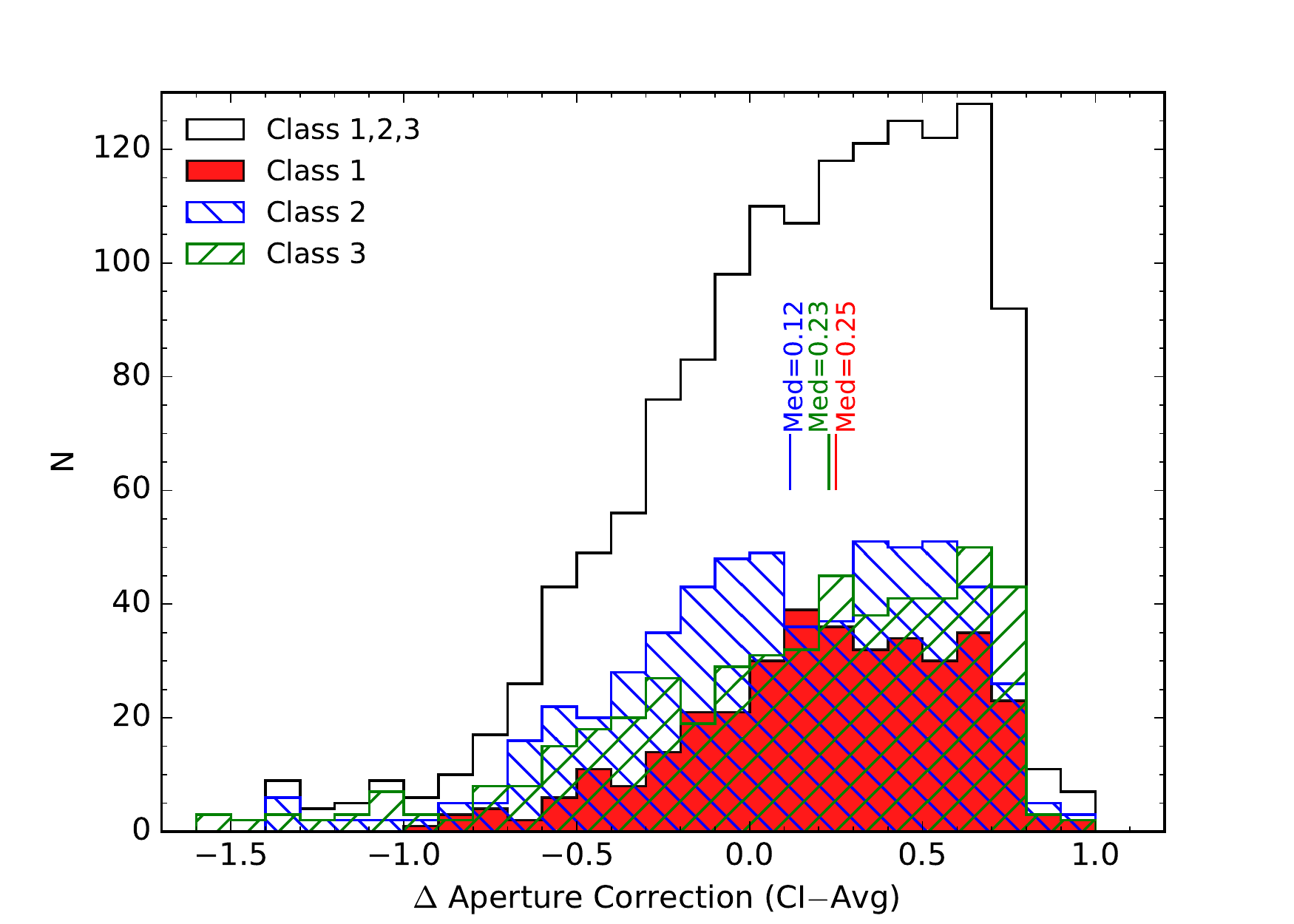}
  \caption{The CI minus the average aperture correction histogram for all clusters in the LEGUS dwarfs as measured in the $V-$band. The distribution shows that the majority of clusters are more compact than the average while there exists a small tail of more extended clusters. The histograms are further broken down into the class 1, 2, and 3 as red-filled, blue line-filled, and green line-filled histograms, respectively. }
   \label{fig:apcorrhist}
   \end{center}
\end{figure}

We first examine how the total cluster fluxes compare given the two aperture corrections studied here. Figure~\ref{fig:apcorrhist} plots the distribution of the differences between the CI and average aperture corrections in the $V$-band for the aggregate dwarf galaxy cluster sample, where the different histograms represent different cluster classes. As shown in  Figure~\ref{fig:apcorrhist} (and previously in Figure~\ref{fig:ciapcorrreal}), corrections inferred from the CI can differ from the average by as much as $\approx$1~mag.

Figure~\ref{fig:apcorrhist} illustrates how common are the extreme extended/compact clusters. Since the aperture corrections are in magnitudes in this figure, we note that objects with negative difference aperture corrections (to the left) have greater CI-based aperture corrections than the average and are thus more extended sources. Conversely, more compact sources will have positive values (to the right) in Figure~\ref{fig:apcorrhist}.


Overall the distributions are not centered on zero, but are shifted to positive values (i.e. more clusters tend to have smaller CI-based aperture corrections) and thus are more compact relative to the isolated training clusters. The median difference for all clusters is 0.2, and the differences for class 1, 2, and 3 are similar.  As might be expected, there is a larger tail of negative values for the class 2 and 3 clusters, indicating that these classes include objects with more extended profiles relative to the isolated training clusters.


The Class 3 sources show a large number of more compact objects; however, there still exists a significant number of extended class 3 sources. Class 3 objects are defined by the groupings of stars (i.e., multiple radial profile peaks), but they exhibit a wide range in morphology and density of these stellar groupings. The more extended Class 3s exhibit a more pronounced unresolved component, while the compact Class 3s tend to have a pronounced stellar object near the center.  



\begin{figure}
  \begin{center}
  \includegraphics[scale=0.5]{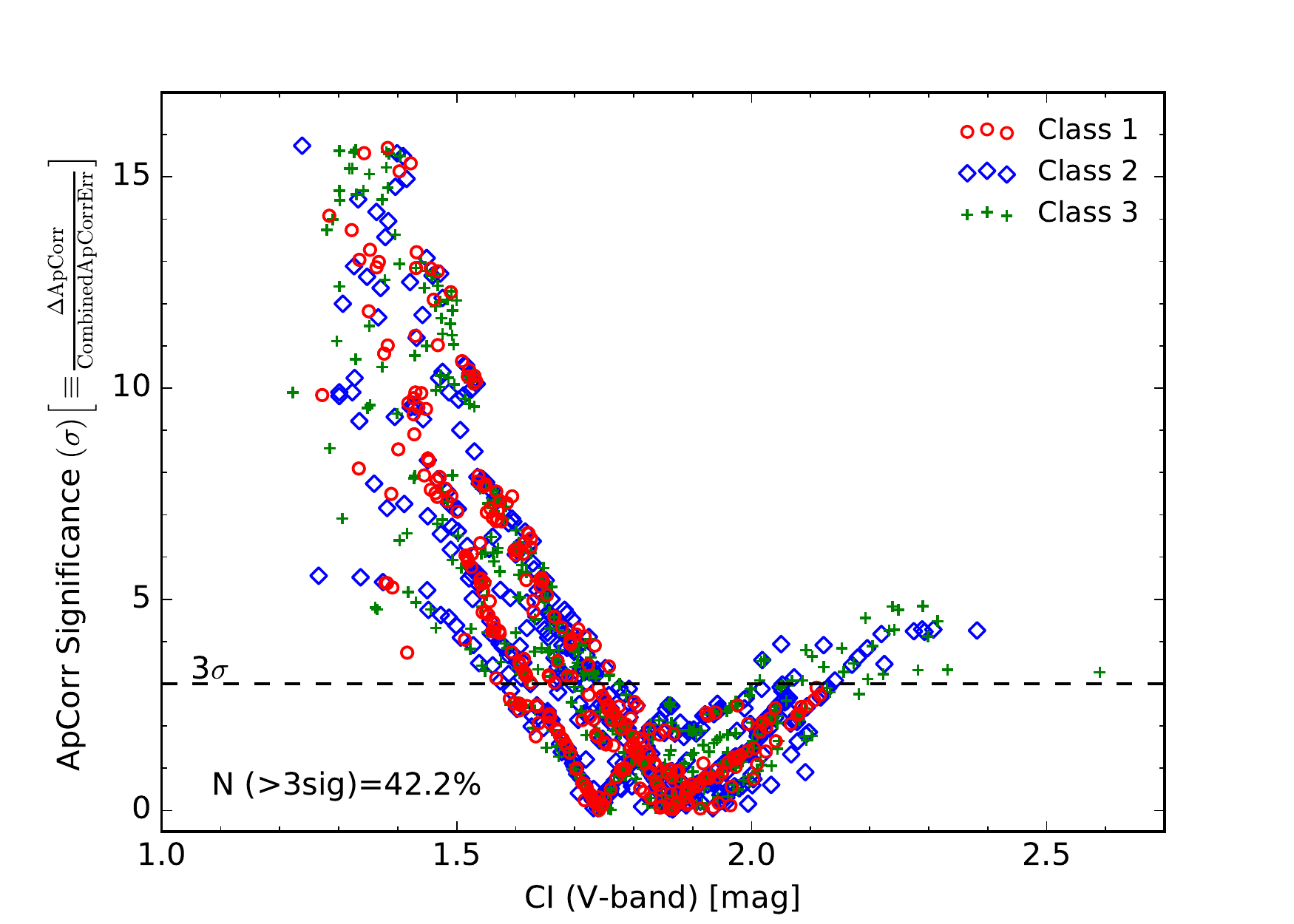}
  \caption{The significance of the difference between the CI and average aperture corrections given their combined measured uncertainties. The significance is defined as the difference in aperture corrections divided by their uncertainties and added in quadrature. Between 40--50\% of all clusters (depending on the filter) show a $>3\sigma$ aperture correction difference indicating that roughly half show a true variation in their radial profile compared to the average training cluster.}
   \label{fig:dapcorrsig}
   \end{center}
\end{figure}  

A natural question to ask is whether the large spread of CI-based aperture corrections is a true reflection of the diversity of radial profiles in the cluster population or whether the spread is mainly due to photometric uncertainties in the measurement of the CI. Thus, we next determine the significance of the differences between the two aperture corrections. In other words, how many sigma ($\sigma$) apart are the two aperture corrections? Figure~\ref{fig:dapcorrsig} presents this significance versus the CI values of all clusters in the $V-$band, which shows that the low (CI$<$1.6) and high (CI$>$2.1) CI clusters are significantly different ($>3\sigma$) from the average. We note that we see similar distributions in all other filters. Between 40--50\% of all clusters in the different filters have significant CI-based aperture corrections compared to the average, and thus are not consistent with the average corrections given their measured errors. \textit{In other words, nearly half of the clusters show true variations in their radial profiles.} Conversely, the other half of the clusters have CI-based corrections which are consistent with the average corrections within the uncertainties and generally have CIs between 1.8 and 2.0~mag.



\begin{figure*}
  \begin{flushleft}
  \includegraphics[scale=1.05]{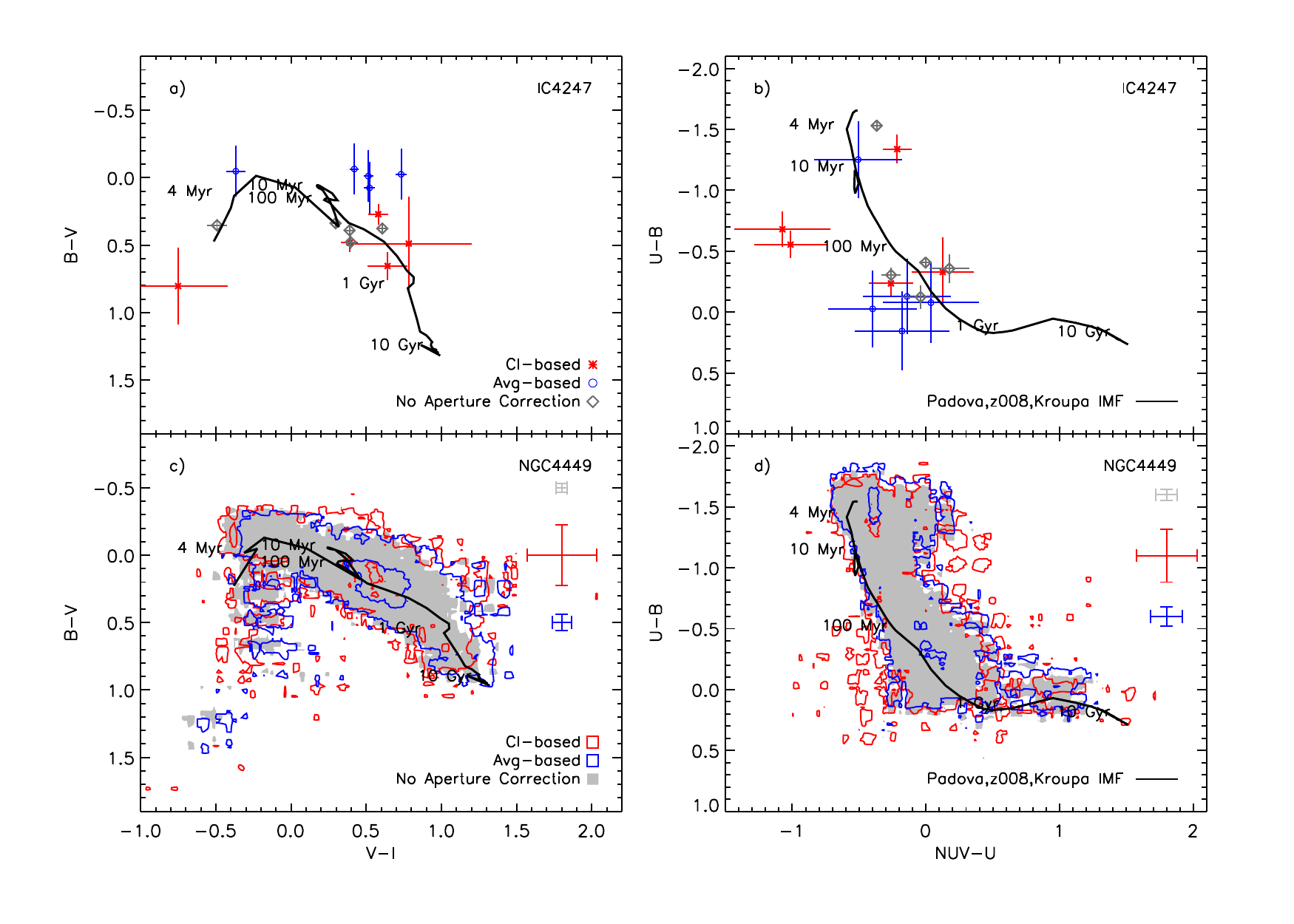}
  \caption{Color-Color plots for both IC4247 (top panels) and NGC4449 (bottom panels), where the solid lines represent the Padova-AGB isochrones with an LMC metallicity and no extinction applied. The left and right panels are a comparison for the two galaxies using the same colors: left: $B-V$ vs $V-I$ and right: $U-B$ vs $NUV-U$. We find that the average (blue circles) and CI-based (red asterisks) aperture corrections show inconsistencies with the model isochrones and colors derived using no aperture correction (gray diamonds) in different situations. The average-based colors are offset for $B-V$ vs $V-I$ by the same amount as the difference in average aperture corrections across filters. However, some CI-based colors disagree with the models and no aperture correction colors in the bluer colors ($NUV-U$) due to marginal detections in these filters. }
   \label{fig:ccplotcomp}
   \end{flushleft}
\end{figure*}  

Next, we examine the impact of the aperture corrections on the cluster colors in the LEGUS dwarfs. Figure~\ref{fig:ccplotcomp} shows different color-color plots for IC4247 and NGC4449. The top panels show IC4247 (a galaxy with only 2 training clusters) while the bottom panels show NGC4449 (a galaxy with many training clusters). For comparison, we also include colors measured with no aperture correction applied, as they may better represent the true cluster colors. We have excluded class 3 sources in Figure~\ref{fig:ccplotcomp} for clarity. 

The left pair of panels show the $B-V$ versus $V-I$ color-color plots for the two galaxies, while the right pair show the $U-B$ versus $NUV-U$ colors.  The colors uncorrected for aperture tend to have the least amount of scatter around the model tracks, while the colors derived from CI-based total fluxes have the most scatter; as might be expected from the relative uncertainties in the colors. Overall, the ensemble populations have consistent color distributions, but clearly the colors can vary significantly for individual clusters. For instance, the average aperture correction colors show systematic offsets with the uncorrected colors, while the CI-based colors tend to show larger scatter in the $NUV$- and $U$-bands. 

In the next section we test how the range in total fluxes and colors between the aperture corrections translate into a difference in the derived physical properties (i.e., age, mass, and extinction).

\subsubsection{Physical Property Comparison} 
The cluster ages and masses are determined via SED fitting to single-aged stellar population models. The methods are detailed fully in \citet{adamo17}, but we provide a brief overview here. The cluster photometry are fit via two methods: 1) with Yggdrasil \citep{zackrisson11} SSP models with the assumption that the IMF is fully sampled and 2) with a Bayesian fitting method based on SLUG \citep[Stochastically Lighting Up Galaxies;][]{slug} where the IMF is stochastically sampled via \textsc{cluster\_slug} \citep{krumholzb,krumholz15}. Since the goal of this paper is to compare the properties of clusters in dwarf galaxies to those in spirals, we have chosen to use the physical properties produced by Yggdrasil methods for a more direct comparison to the results of LEGUS spirals studied in \citet{adamo17} and \citet{messa18a}.

The Yggdrasil method uses the model parameters available in Starburst99 \citep{starburst99,vazquez05}, where two commonly used stellar libraries (Padova-AGB and Geneva tracks) with a Kroupa IMF that ranges from 0.1 to 100~$M_{\odot}$ are provided as well as three extinction laws: Milky Way \citep{cardelli89}, starburst \citep{calzetti00}, and starburst with differential extinction for stars and gas. These models are input into Cloudy \citep{cloudy} to produce fluxes from nebular emission lines and continuum. In this analysis, we use the SED output based on the following assumptions: the Padova-AGB libraries, a starburst extinction law with differential reddening, and the measured gas phase metallicity of each galaxy (see Table~\ref{tab:genprop}). 

\begin{figure}
  \includegraphics[scale=0.48]{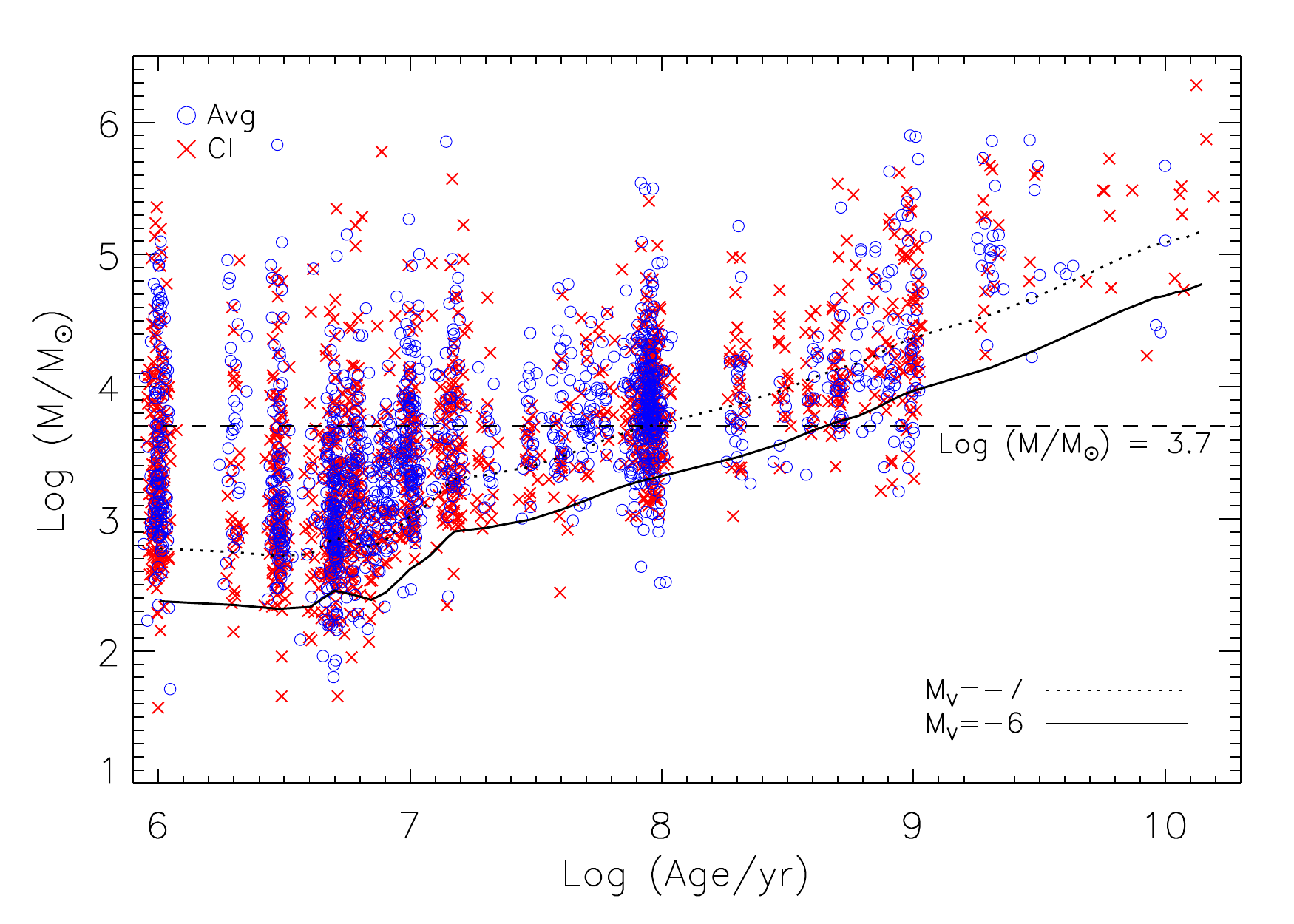}
  \caption{The age-mass diagram for all clusters in the LEGUS dwarf galaxies. The solid line is the Padova isochrone corresponding to the absolute $V-$band magnitude cut of $M_V=-6$~mag.}
   \label{fig:agemass}
\end{figure}  

Figure~\ref{fig:agemass} shows the age-mass diagram for all clusters found in the LEGUS dwarf galaxies for both CI- and average-based photometry. The majority of the clusters below the $M_V{=}-6$~mag cut line are those that have been identified via visual inspection. There is broad agreement between the CI- and average-based aperture corrected fluxes in the coverage of this diagram.

\begin{figure*}
  \includegraphics[scale=0.5]{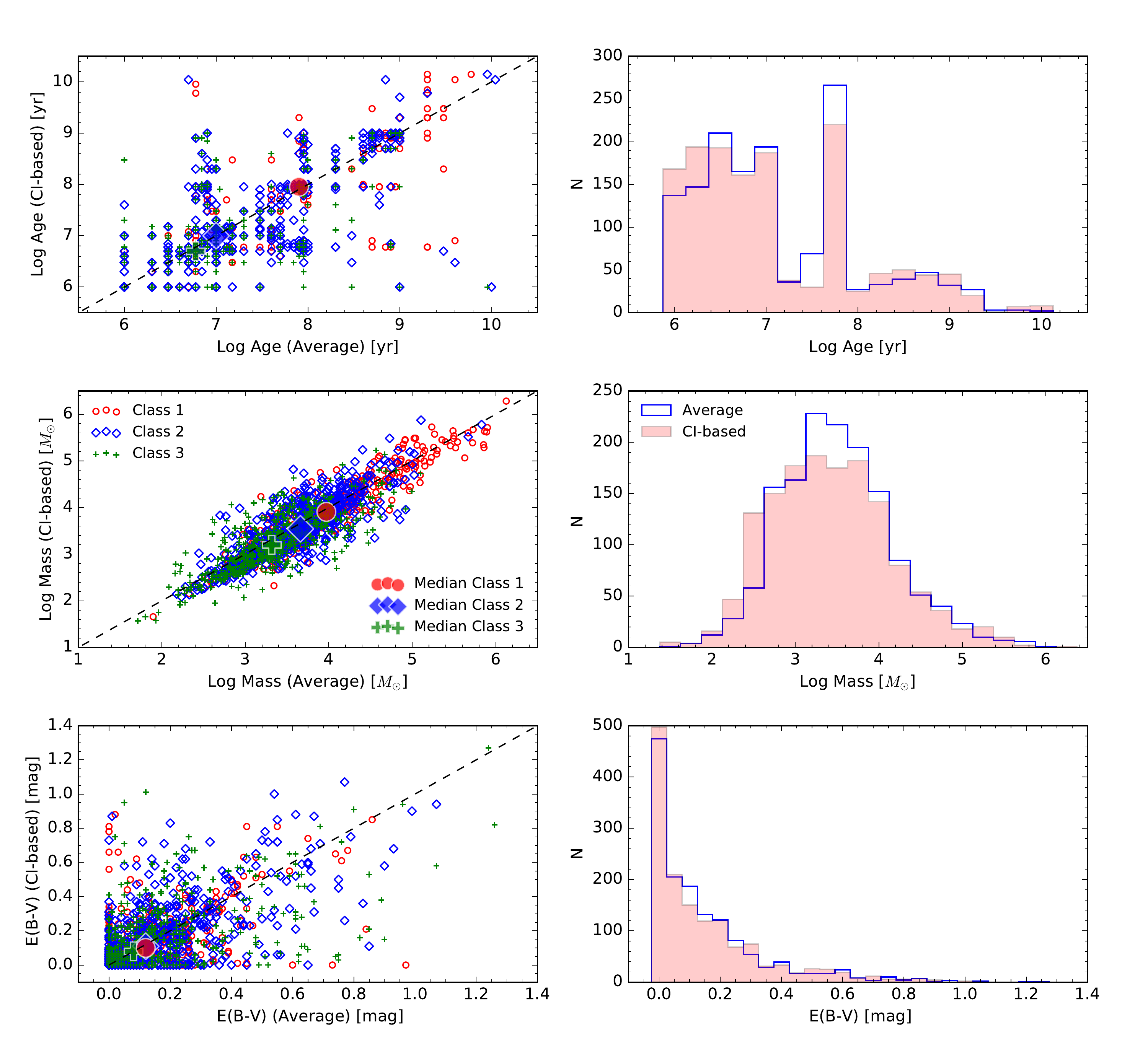}
  \caption{A six panel plot where each row compares the age (top), mass (middle), and extinction (bottom) of all clusters via a 1-to-1 scatter plot (left) and a histogram panel (right). We find overall agreement for the physical properties derived from the average- and CI-based aperture corrections. However, we do find that the CI-based masses are smaller by 0.1~dex. This is likely due to the majority of clusters with a smaller CI-based aperture correction than the average (see the difference aperture correction histograms of Figure~\ref{fig:apcorrhist}). }
   \label{fig:agemasscomp}
\end{figure*}  

Figure~\ref{fig:agemasscomp} is a six panel plot comparing ages (top), masses (middle), and extinctions (bottom) derived from the CI and average aperture corrected fluxes for all clusters across the LEGUS dwarf galaxy sample. Scatter plots are shown on the left, and histograms are shown on the right. A comparison of ages shows overall agreement, but with large scatter.  We find a median difference of 0.0 with a standard deviation of 0.64~dex for all cluster classes. The age histograms show a similar distribution. This is similar to what \citet{adamo17} found for the LEGUS spiral galaxy NGC628. We note that the apparent age gap between 7.2 - 7.6 is a well known artifact \citep{maiz09} and does not imply a real deficit in this age range. This feature arises because the models loop back on themselves during this time period, covering a fairly large range in age but a small range of colors. This also explains the pile-up of clusters at 7.8 in a log(age).


A comparison of masses shows overall agreement with a smaller degree of scatter (0.37~dex). However, the masses derived from the CI-based aperture corrections are systematically lower where the median difference is 0.1 dex. This small overall shift to lower masses can also be seen in the histograms. The shift in masses can be understood from inspection of Figure~\ref{fig:apcorrhist}, where we found that the median CI-based aperture correction for all clusters was 0.2~mag fainter than those derived from the average-based aperture corrections. 

A comparison of extinction values shows overall agreement for the majority of clusters with some scatter, where the median difference is 0.0 with a standard deviation of 0.18~dex. The histogram comparisons also show little difference between the derived extinctions for the CI- and average-based aperture corrections. We note that the median extinction for all clusters in the LEGUS dwarf galaxies is 0.1~mag which suggests that these dwarf galaxies have low extinction environments \citep{lee09b,hao11,kahre18}. 

As can be seen in the scatter plots, the age and mass distributions are different for the Class 1, 2, and 3 clusters \citep{grasha15,grasha17,adamo17}.
We find that class 1, 2, and 3 sources have a median log(age,yr) of 8.0, 7.0, and 6.7, respectively; these values are similar for ages derived using both aperture corrections. The trend between age and class suggests that the associations (Class 3) are the youngest population while the more compact population (Class 1) is the oldest. We also find that Class 1, 2, and 3 have a median log(mass,M$_{\odot}$) of 3.9, 3.6, and 3.2, respectively. Thus, the Class 1 clusters are the oldest and most massive, while the associations (Class 3) are the youngest and least massive.




\begin{figure}
  \includegraphics[scale=0.53]{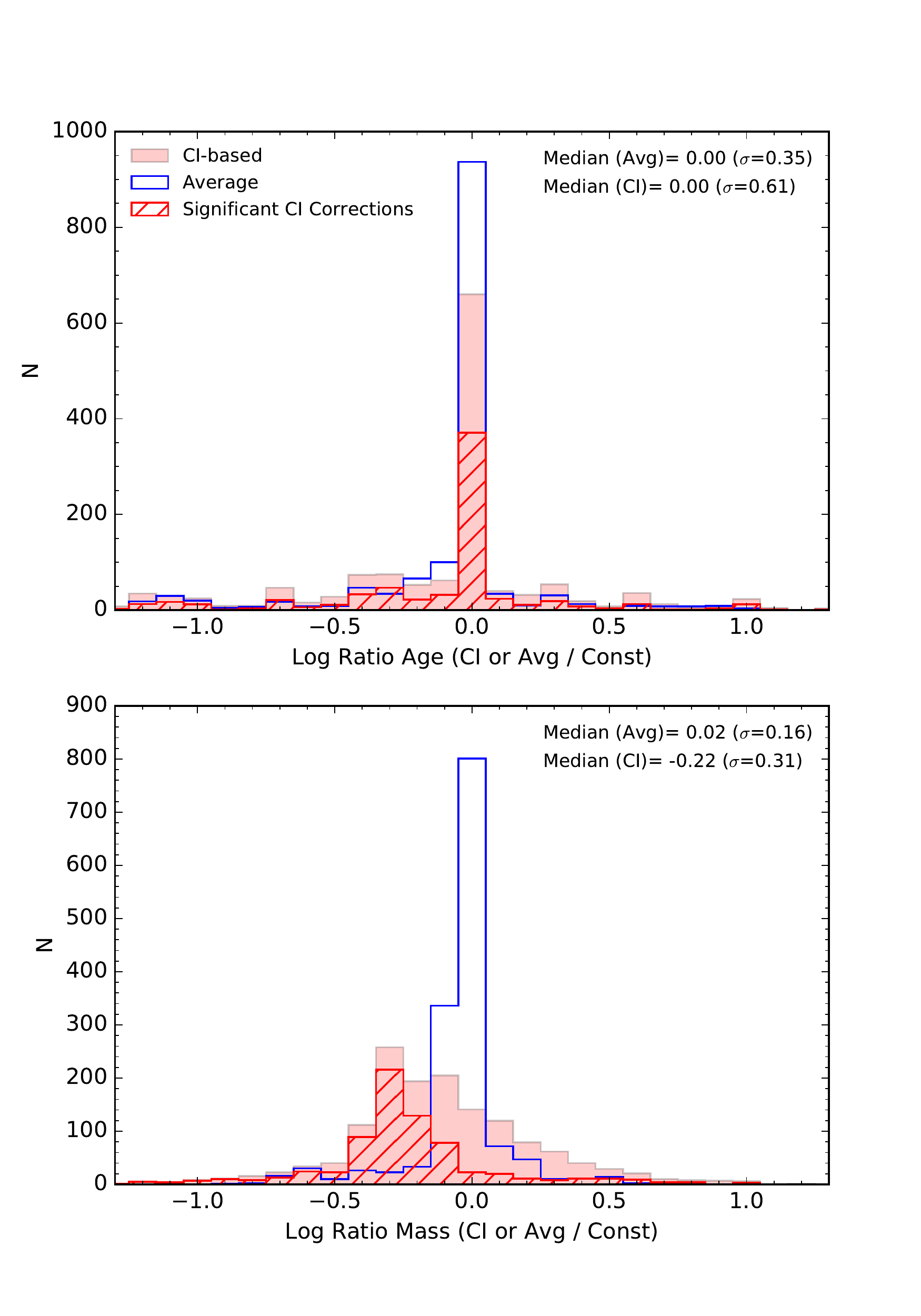}
  \caption{The age (top panel) and mass (bottom panel) ratios for the average- or CI-to-constant aperture correction as represented by open blue and filled red histograms, respectively. The constant aperture correction is defined as the median of all average corrections with a value of --0.85~mag. The hashed red histograms represent the clusters whose CI-average aperture correction differences are greater than 3$\sigma$ of the combined correction errors. The average ages and masses show good agreement with the constant correction ages and masses. The CI ages agree with the constant ages with increased scatter, but the CI masses are 0.2~dex smaller than those derived with a constant correction. The majority of the clusters with low CI-to-constant mass ratios are those with a significantly different CI-based aperture correction suggesting that many of the lower CI-based masses reflect a real difference in cluster mass.}
   \label{fig:agemass4px}
\end{figure}  


To further explore how our aperture corrections can affect the derived ages and masses, we re-compute them using fluxes where a single, constant aperture correction of 0.85~mag has been applied across all filters for all clusters (i.e., the median average aperture correction for all filters in all dwarf galaxies). This process leaves the colors unchanged, and provides a useful comparison as they may better represent the true cluster colors as illustrated in Figure~\ref{fig:ccplotcomp}.  In Figure~\ref{fig:agemass4px} we show histograms for the ratios of ages and masses computed using the CI or average aperture corrected fluxes relative to those computed with a 0.85~mag constant correction.  As might be expected based upon the color-color diagrams and model tracks shown in Figure~\ref{fig:ccplotcomp}, the ages are in overall agreement (the distribution of age ratios are centered upon a value of unity), but show large scatter (factor of $>$2).  The ages derived from the CI-based aperture corrected fluxes show a larger spread in values.  

Examination of the mass histograms show broad agreement between those based on the constant and average aperture corrected fluxes, but there exists an offset between those based on the constant- and CI-based aperture correction (the median ratio is --0.22~dex).  This is a consequence of the fact that many of the clusters are more compact compared to the training clusters, and thus have total fluxes that are overestimated by the average (and similarly the constant) aperture correction. In addition, a large fraction of clusters with low CI-to-constant mass ratios are dominated by those with a statistically significant CI minus average correction difference (as represented by the hashed histogram). Thus, the lower CI-based masses likely reflect a real difference in the derived masses from either the average or the constant aperture correction.

\subsubsection{Cluster Distribution Functions (Age, Luminosity, and Mass)} \label{sec:distfunc}

In this section we explore the effects of our two aperture corrections on the distribution functions of age, luminosity, and mass in the LEGUS dwarf galaxy clusters. Here we focus on class 1 and 2 clusters; however, we note that we find similar results when including the class 3 sources. 

\begin{figure}
  \includegraphics[scale=0.5]{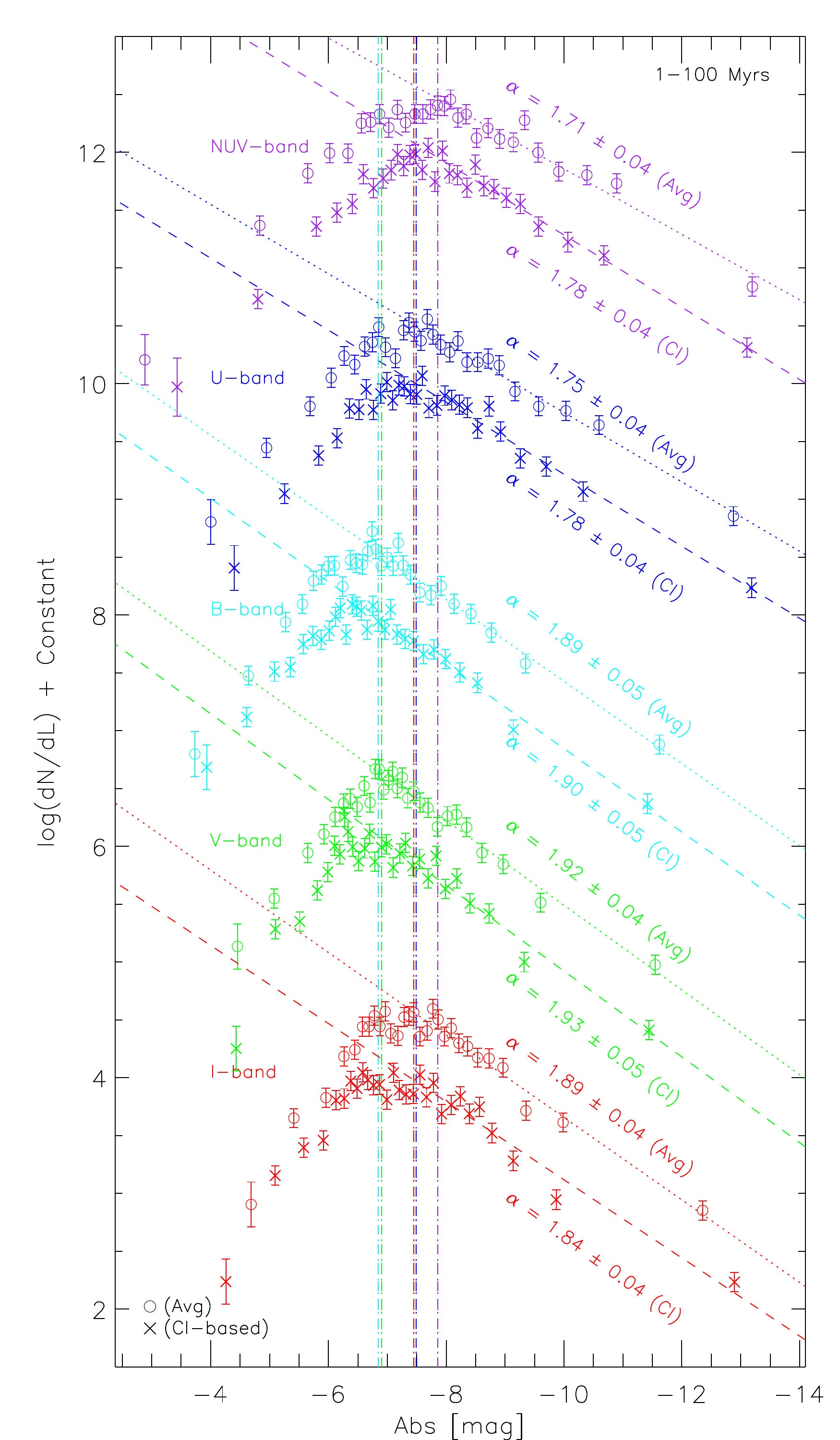}
  \caption{The luminosity functions (LFs) for all clusters with an age less than 100~Myr in all LEGUS dwarf galaxies. The LF slopes show agreement across all filters within the errors. However, the bluer filter LF slopes tend to be flatter than the redder wavelength LF slopes.}
   \label{fig:LFci}
\end{figure}

Figure~\ref{fig:LFci} shows the LFs for the clusters in all LEGUS dwarfs with an age less than 100~Myr using both CI- and average-based aperture corrections. Each filter's LF is color coded and the y-axis has been normalized to an arbitrary number for clarity. The binned LFs have been constructed with an equal number of clusters in each luminosity bin \citep{miaz05} where the y-axis is calculated as the number of clusters per bin divided by the bin width. For more details on the constructing these distributions see \S5.1 of \citet{cook16}. We derive the LF slope via fitting a power-law to bins with luminosities brighter than the peak of the luminosity histogram \citep[][]{cook16}. We note that the peak of the luminosity histogram agrees with the turnover found in the LFs for each filter. 

A comparison of the LF slopes between the CI- and average-based aperture corrections reveals no difference within the fitted errors for each of the five filters. However, we do find that the bluer ($NUV$ and $U$) LF slopes tend to be flatter than those at longer wavelengths ($BVI$) as was found by other studies of spiral galaxies \citep{DolphinKennicut02,elmegreen02,gieles06a,haas08,cantiello09,gieles10,chandar10b,adamo17}. The median $NUV-$ and $U-$band slopes are 2.8$\sigma$ flatter than the median $BVI-$band slopes. We note that we find similar results when using more conservative magnitude cuts (a few tenths of a magnitude) than the peak when fitting the LF slopes.


\begin{figure}
  \includegraphics[scale=0.5]{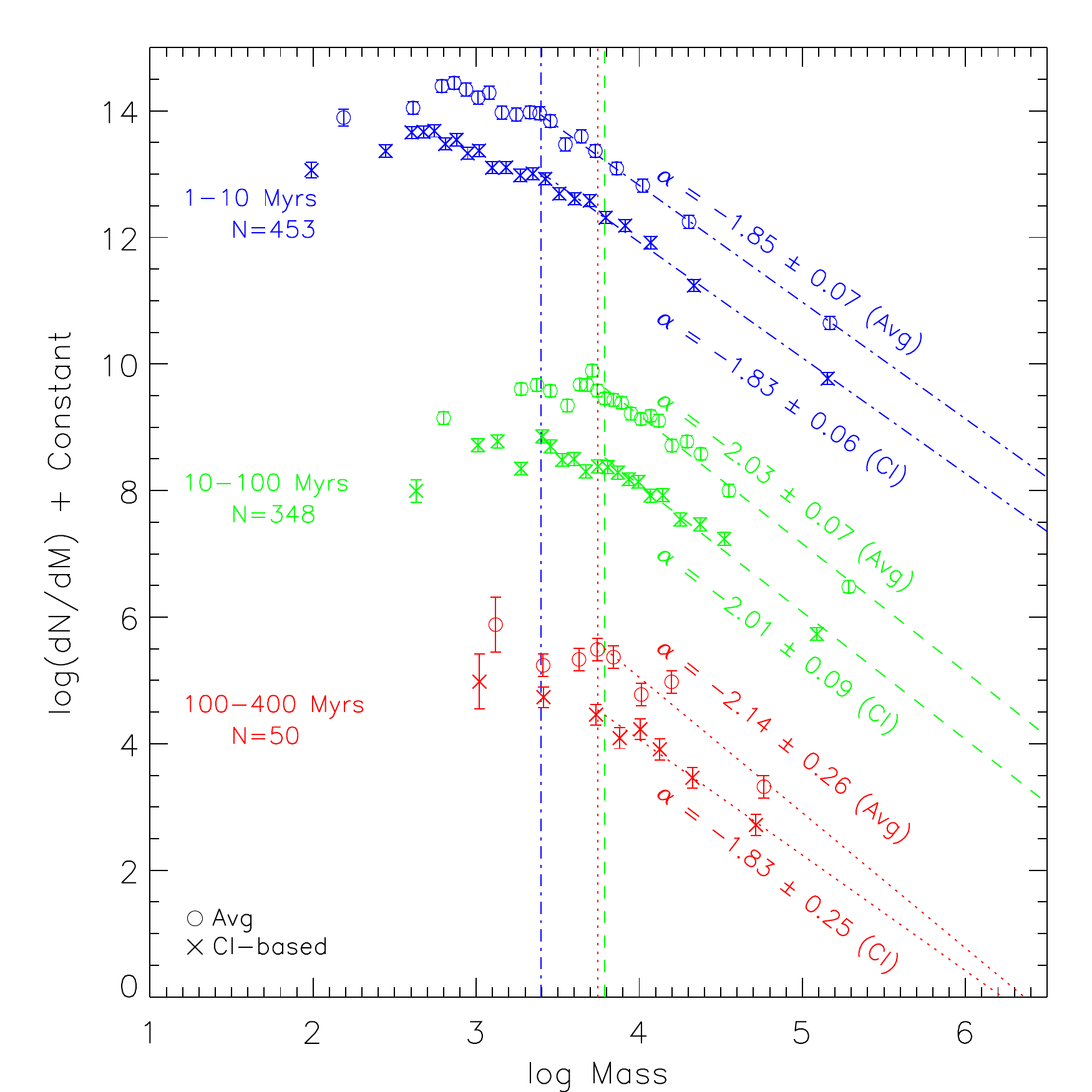}
  \caption{The mass functions (MFs) for all clusters in the LEGUS dwarf galaxies broken into three age ranges: 1-10~Myr, 10-100~Myr, and 100-400~Myr. The MF slopes for all three age bins agree with the canonical $-2$ power-law slope. }
   \label{fig:MFci}
\end{figure}  

Figure~\ref{fig:MFci} shows the cluster MFs in different age ranges (1-10, 10-100, and 100-400~Myr) using both CI- and average-based aperture corrections. We find no difference in the MF slopes between the two aperture corrections for all three age ranges given the uncertainties. We also find a similar MF slope for all three age bins of $-1.9\pm0.1$, which is consistent with a canonical $-2$ power-law slope found by many previous clusters studies \citep{battinelli94,elmegreen97,zhang99,hunter03,bik03,degrijs03b,mccrady07,chandar10b,cook12,adamo17}. 


\begin{figure}
  \includegraphics[scale=0.5]{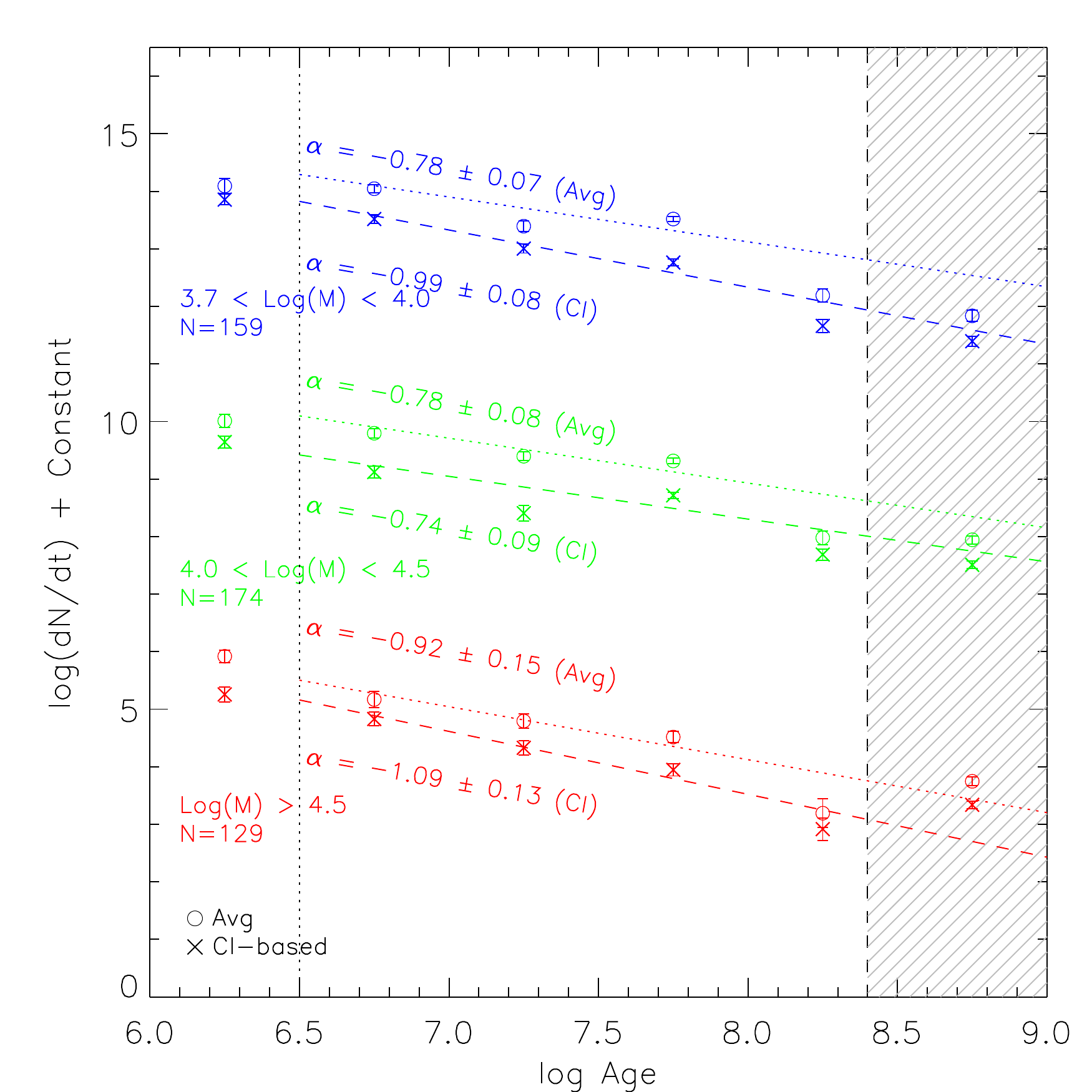}
  \caption{The age distributions of all clusters in the LEGUS dwarfs broken into three logarithmic mass bins: 3.7-4.0, 4.0-4.5, and $>$4.5. We make a mass cut above log(mass) of 3.7 to avoid incompleteness at older ages and to avoid variations due to stochastic IMF sampling of low-mass clusters. Following \citet{adamo17} and \citet{messa18a}, we exclude the youngest age bin. We find a power-law slope of $-0.8\pm0.15$ for all mass bins and for both aperture corrections. }
   \label{fig:dndtall}
\end{figure}

Figure~\ref{fig:dndtall} shows the cluster age distributions in different logarithmic mass bins (3.7$-$4.0, 4.0$-$4.5, and $>$4.5) using both CI- and average-based aperture corrections. We use a log(mass/M$_{\odot}$) cut of 3.7 to avoid incompleteness at older ages and to avoid variations in the derived physical properties of clusters due to stochastic sampling of the cluster IMF \citep{fouesneau10,krumholz15}. Note that we use a mass cut instead of a luminosity cut since the derived masses will have taken into account the fading of clusters over time \citep{fall05}. We exclude the youngest age bin following the methodology of \citet{adamo17} and \citet{messa18a}. However, we note that the youngest age bin data agree with the fitted distribution at older ages. We find no significant difference between the average- and CI-based age distributions. In addition, we find no difference across the populations in mass bins, and find a median $-0.8\pm0.15$ power-law slope for all three mass bins.


It is possible that our choice of age bin size and age range might affect our fitted age distribution slopes. As such, we have tested the bin sizes and age ranges used in our age distribution fits. Neither smaller nor larger bin sizes show significant differences in their distribution slopes, but we do find larger fluctuations in the smallest bin sizes ($\Delta t$=0.2) most of which reflect the known age gap artifacts due to model fitting. We also test fitting a power law to ages above 10$^7$~yrs, and find a steeper slope of --1.1, but find no difference between the aperture corrections nor across the mass bins. It is also possible that young bursts of star formation in some of our dwarfs with N$>$100 clusters (NGC4449, NGC4656, and NGC3738) may dominate the age distributions and artificially create a steeper slope. We test this by removing each and all of these three galaxies and find no differences in the age distribution slopes within the errors.


{\it We conclude that any discrepancies found in the total fluxes, colors, ages, and masses of individual clusters when using different aperture corrections do not translate into a measurable effect in the LFs, MFs, nor the age distributions of ensembles of star clusters. }




\subsubsection{Aperture Correction Comparison Take-Away Points}
We have performed an analysis regarding the effects of two commonly-used cluster aperture corrections on both the observable and physical properties of star clusters. Both methods show consistent luminosity, mass, and age distributions for ensembles of clusters, but both have drawbacks when measuring the properties of individual clusters.

The average aperture correction can produce systematic color offsets when too few training clusters ($N<10$) are available to define the average correction. In addition, the CI-based aperture corrections show increased color scatter for clusters with marginal detections in some filters (usually the $NUV$- or $U$-bands). The ages from both the CI- and average-based aperture corrections show larger scatter compared to those derived from a constant correction, where the CI-based ages ($\sigma$(CI-to-constant)=0.6) are larger than the average ($\sigma$(Avg-to-constant)=0.35~dex).



The median relative difference in total flux resulting from the two aperture corrections is 0.2 mag (in the sense that the average correction is larger indicating that most of the clusters are more compact than the training clusters), and that the difference for individual clusters can be as large as $\sim$1~mag (Figure~\ref{fig:apcorrhist}).  For half of the clusters in our dwarf galaxy sample, these differences are within the photometric uncertainties; for the other half the difference points to a true variation in the radial profile of the clusters relative to those characterized by the training sample. The median difference in total fluxes translates into a median mass difference of 0.22 dex where the masses derived from CI-based corrections are smaller than the average.


From these experiments we have found that the total fluxes of individual clusters are more accurately recovered from a CI-based aperture, but that the CI-based aperture corrections result in increased scatter around the predicted colors when applied individually to each filter. Based on these results, we recommend the following hybrid strategy for aperture corrections. Measure the CI using the filter in which the clusters are detected (V-band in this case), assume the same CI for all other bands, and compute the appropriate aperture correction given the HST camera (i.e., the appropriate aperture correction-CI relationship from Table~\ref{tab:fakeclust}). This method will introduce a small amount of scatter in the final fluxes across filters due to the PSF variation across the two HST cameras, but this scatter will be smaller than the scatter added by either aperture correction studied here. We have implemented these recommendations on a single galaxy NGC 4449 with significant clusters and find similar age and mass distributions within the fitted errors.





\subsection{Comparison to ANGST Dwarf Galaxy Clusters \label{sec:angstcomp}}

In this section we compare the cluster populations found in the LEGUS dwarf galaxies to those found in the ANGST dwarf galaxies \citep{cook12}. The cluster catalogs in these two programs represent two of the largest dwarf galaxy samples to have uniformly identified and characterized clusters. However, these two programs use two different identification methods. Thus, a comparison of their cluster populations can yield insights into effective identification methods in these extreme environments. 

The main difference in cluster identification methods between ANGST and LEGUS is the generation of star cluster candidates. The ANGST cluster candidates were identified via visual inspection of HST images whereas the LEGUS cluster candidates were generated via automated methods. Both programs then used visual classification, with similar classification definitions, to produce final cluster catalogs. 

\input{tables/clustnum.tex}

The ANGST dwarf sample consisted of 37 galaxies whose global SFRs extended down to log(SFR) of --5 where 144 clusters were found at all ages. There are three galaxies in common between ANGST and LEGUS: UGC4305, UGC4459, and UGC5139. In these three galaxies LEGUS found 2.5 and 11 times the number of clusters found in ANGST for all ages and $<$100~Myr, respectively (see Table~\ref{tab:clustnum}). However, it should be noted that the LEGUS pipeline produces many more candidates that are rejected by visual classification. 


\begin{figure}
  \begin{center}
  \includegraphics[scale=0.26]{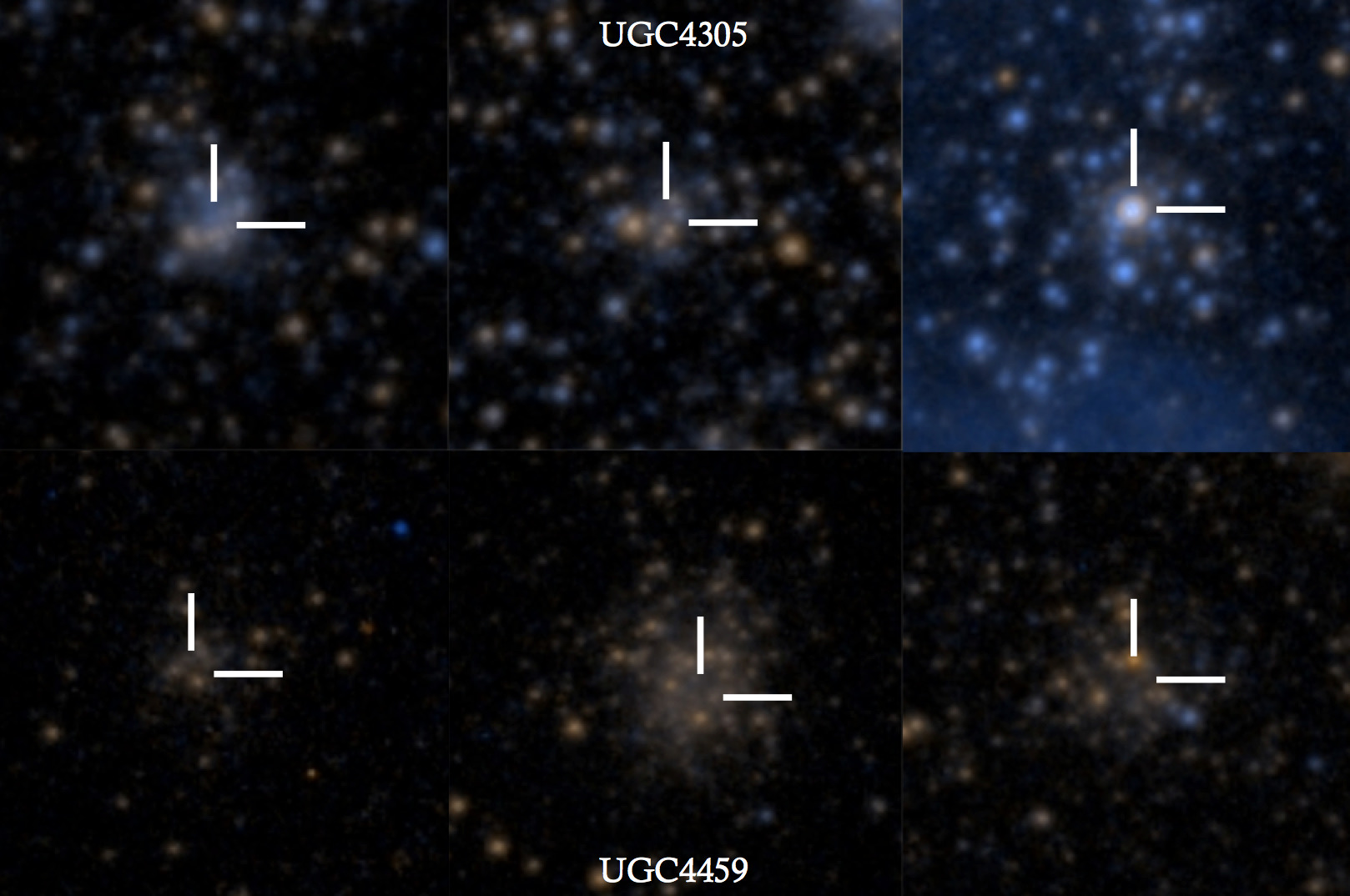}
  \caption{Six ANGST clusters missed by the LEGUS identification methods, where 3 are from UGC4305 (top) and 3 are from UGC4459 (bottom). All but one of these sources (except the top-right) are fainter than the $M_V{=}-6$ magnitude cut employed by the LEGUS cluster pipeline. The source in the top-right exhibits a stellar CI value and is likely a bright star on top of a stellar field in the galaxy. The size of the bars are $\sim$0.8\farcs~Five of these were older than 100~Myrs and the upper right was 15Myr.}
   \label{fig:legusmissed}
   \end{center}
\end{figure}

A cross-match of the clusters in both programs shows that all but six of the ANGST clusters were found in the LEGUS catalog. Figure~\ref{fig:legusmissed} shows HST color cutouts of these clusters where 5 are fainter than the LEGUS pipeline magnitude cut ($M_V{=}-6$) and the sixth (upper-right object) exhibits a small CI value consistent with a stellar PSF. Thus, these clusters do not make either the magnitude cut nor the CI cut of the LEGUS pipeline. However, we note that these ``missed" clusters would be visually classified as class 3s or a contaminant in the LEGUS classification scheme. 

\begin{figure}
  \begin{center}
  \includegraphics[scale=0.55]{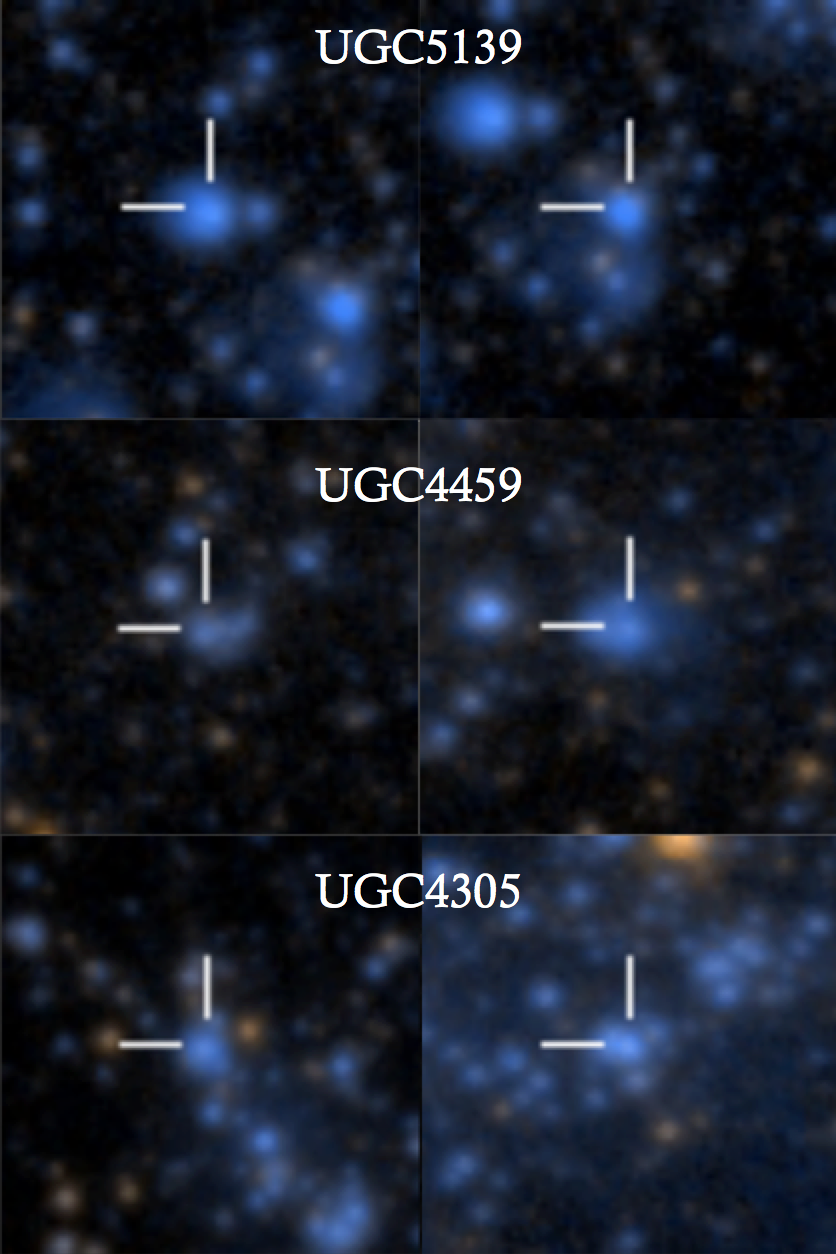}
  \caption{Two-color images of six representative LEGUS clusters that are not in the ANGST cluster catalog (2 from each of the three galaxies). Nearly all of the LEGUS clusters not in the ANGST catalog are compact sources with CI values near 1.6~mag.}
   \label{fig:angstmissed}
   \end{center}
\end{figure}  

\begin{figure}
  \includegraphics[scale=0.5]{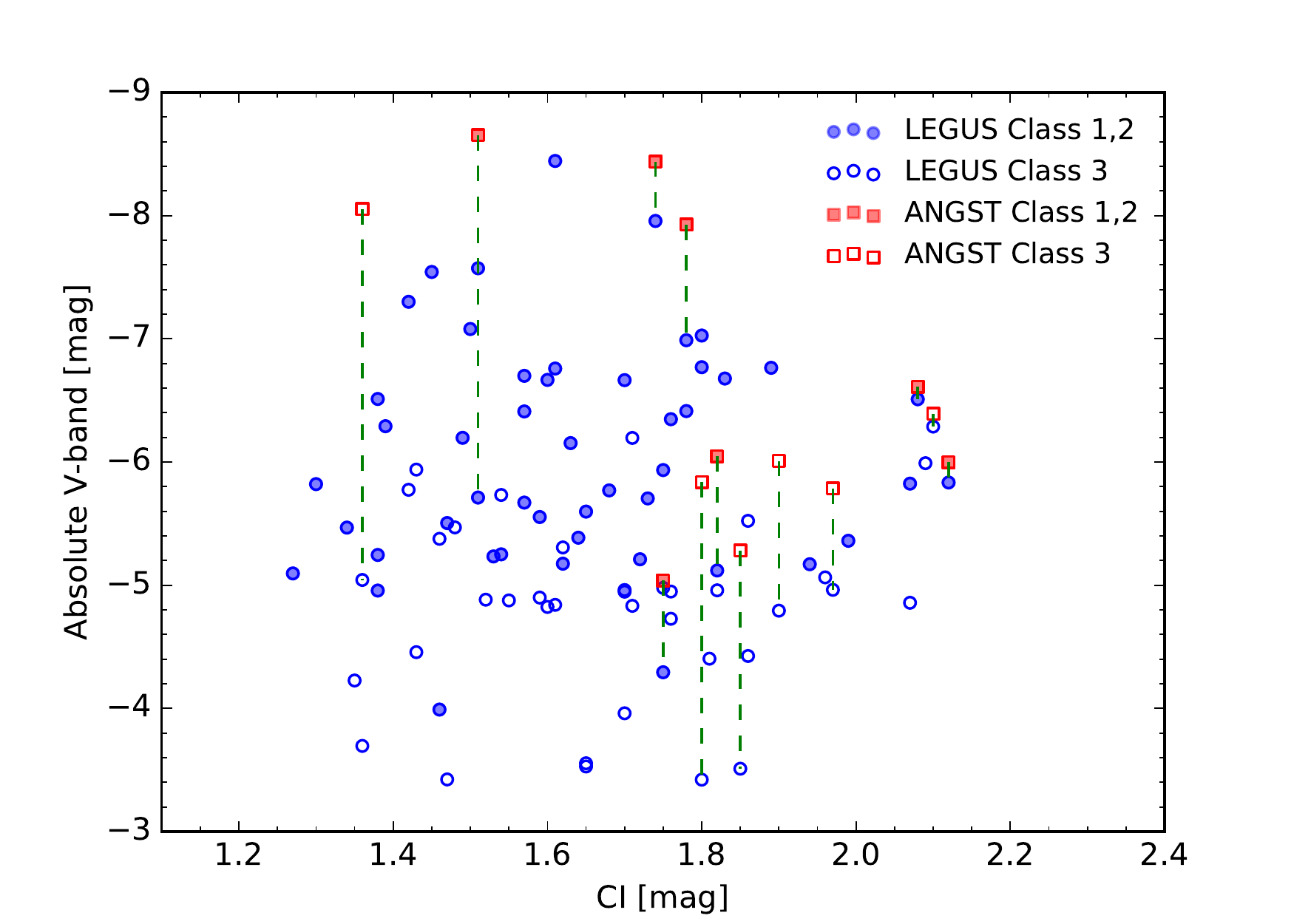}
  \caption{The absolute $V-$band magnitude versus CI for the young ($<$100~Myr) ANGST and LEGUS clusters in the three galaxies, where open symbols represent class 3 sources and closed symbols represent class 1 and 2 clusters. The vertical lines connect the $V-$band magnitude for the same cluster in ANGST and LEGUS. }
   \label{fig:cimagangst}
\end{figure}

The LEGUS pipeline found over twice as many clusters in these three galaxies. We show 2 representative LEGUS clusters missed by ANGST in Figure~\ref{fig:angstmissed} for each of the three galaxies, where these clusters tend to be more compact (CI$<$1.7~mag). To illustrate this, Figure~\ref{fig:cimagangst} shows the absolute $V-$band magnitude versus CI for the young clusters in both ANGST and LEGUS. We use the LEGUS class and CI values for the ANGST clusters. The majority of the LEGUS clusters missed by ANGST tend to be fainter and more compact, which could be difficult to separate from stars in dense regions via visual inspection.


Figure~\ref{fig:cimagangst} also shows that the total flux is overestimated in the ANGST catalog. This is due to the existence of only two high-resolution HST images at the time of the ANGST cluster study \citep{cook12}. Thus, ground-based imaging was used to fill in the wavelength gaps in the cluster SEDs, and the HST imaging was smoothed to match the ground-based seeing. The photometric aperture used by \citet{cook12} was 2$\farcs$5 which is $\sim$10 times the size of the LEGUS photometric aperture. Consequently, there can be considerable contamination from nearby sources in the ANGST photometric aperture as evidenced by several ANGST clusters showing $>$1~mag brighter than the LEGUS CI-based photometry. A comparison of the clusters in common to both the ANGST and LEGUS catalogs, the ages show good agreement. However, the ANGST masses can be significantly larger since derived masses will scale with the measured brightness, where on average the mass ratio is a factor of a few. We note that the clusters with the largest magnitude difference ($\sim3$~mag) have mass estimates in ANGST that are larger by factors of $\sim10-50$.

\begin{figure}
  \includegraphics[scale=0.5]{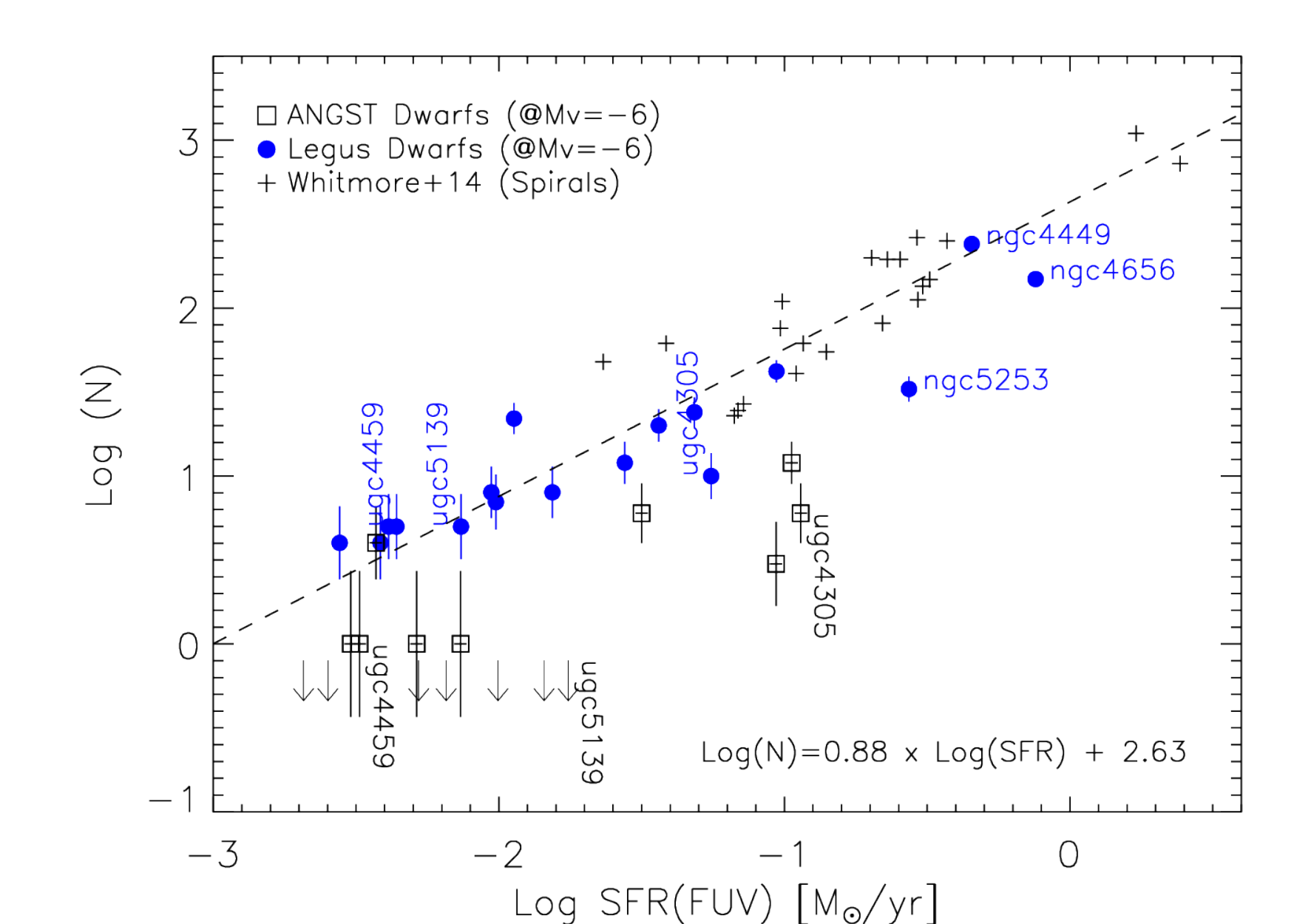}
  \caption{The number of clusters versus the global galaxy SFR for the LEGUS clusters (blue circles), ANGST clusters (open squares), and a uniformly identified catalog of clusters in several spiral galaxies \citep[crosses;][]{whitmore14}. The dashed line is a bisector fit to the LEGUS and spiral sample with an RMS scatter of 0.24 dex.}
   \label{fig:Nsfr}
\end{figure}

To put the cluster statistics of both ANGST and LEGUS into perspective, Figure~\ref{fig:Nsfr} shows the total number of young clusters ($<$100~Myr) found in LEGUS and ANGST given an absolute magnitude cut of $M_V=-6$~mag (i.e., the LEGUS pipeline cut). We note that both studies used the same HST images to identify clusters, thus applying the same magnitude cut to ANGST is reasonable. The ANGST clusters show a consistent dearth of clusters at nearly all SFRs compared to LEGUS. For a non-dwarf comparison, we overplot the number of clusters found in a sample of spiral galaxies brighter than the adopted brightness limits of each galaxy \cite[typically $-8$~mag;][]{whitmore14}, and find that the LEGUS cluster numbers smoothly extend the spiral relationship between the number of clusters and global SFR.

\subsection{Trends With Global Galaxy Properties \label{sec:trends}}


\subsubsection{Binned Age, Luminosity, and Mass Distributions}

Here we explore how the age, luminosity, and mass distributions of clusters in the LEGUS dwarf galaxies change as a function of global galaxy SFR. While most of the LEGUS dwarf galaxies do not have enough young clusters to provide well-behaved distribution functions, we can combine all of the clusters from these galaxies to make a composite dwarf and improve our cluster statistics. This approach has been used by several previous cluster studies \citep{cook12,whitmore14,cook16}.

\begin{figure}
  \begin{center}
  \includegraphics[scale=0.5]{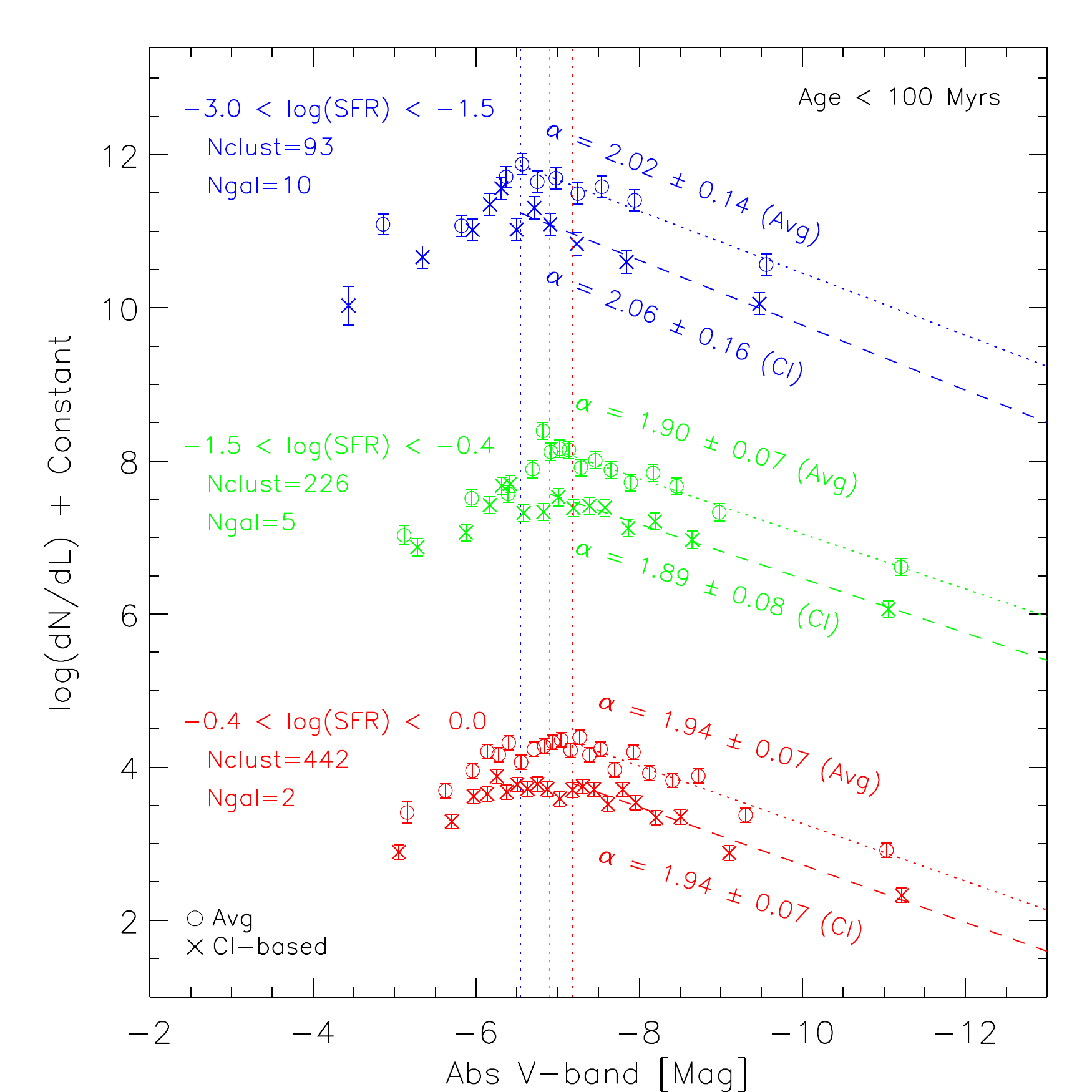}
  \caption{The luminosity functions (LFs) for the clusters in all LEGUS dwarf galaxies binned by global SFR. We find no trend between the LF slope and binned SFR. }
   \label{fig:binLF}
   \end{center}
\end{figure}



Figure~\ref{fig:binLF} shows the LFs of all young clusters (age$<$100~Myr) in the LEGUS dwarfs binned by their host-galaxy's SFR. The SFR bins were chosen to ensure good number statistics in the bins and so that at least two galaxies were in each bin. We find no trend between the binned LF slope and global SFR where the median LF slope is --1.95$\pm$0.06. We also tested various SFR bin definitions and using a color cut to approximate a 100~Myr age cut ($U-B<-0.5$~mag). We found no significant differences in the LF slopes. We note that the luminosity of the brightest luminosity bin increases with the binned SFR as would be expected since the brightest cluster in a galaxy scales with the global SFR \citep{whitmore00,larsen02,bastian08,cook12,whitmore14}.

\begin{figure}
  \begin{center}
  \includegraphics[scale=0.5]{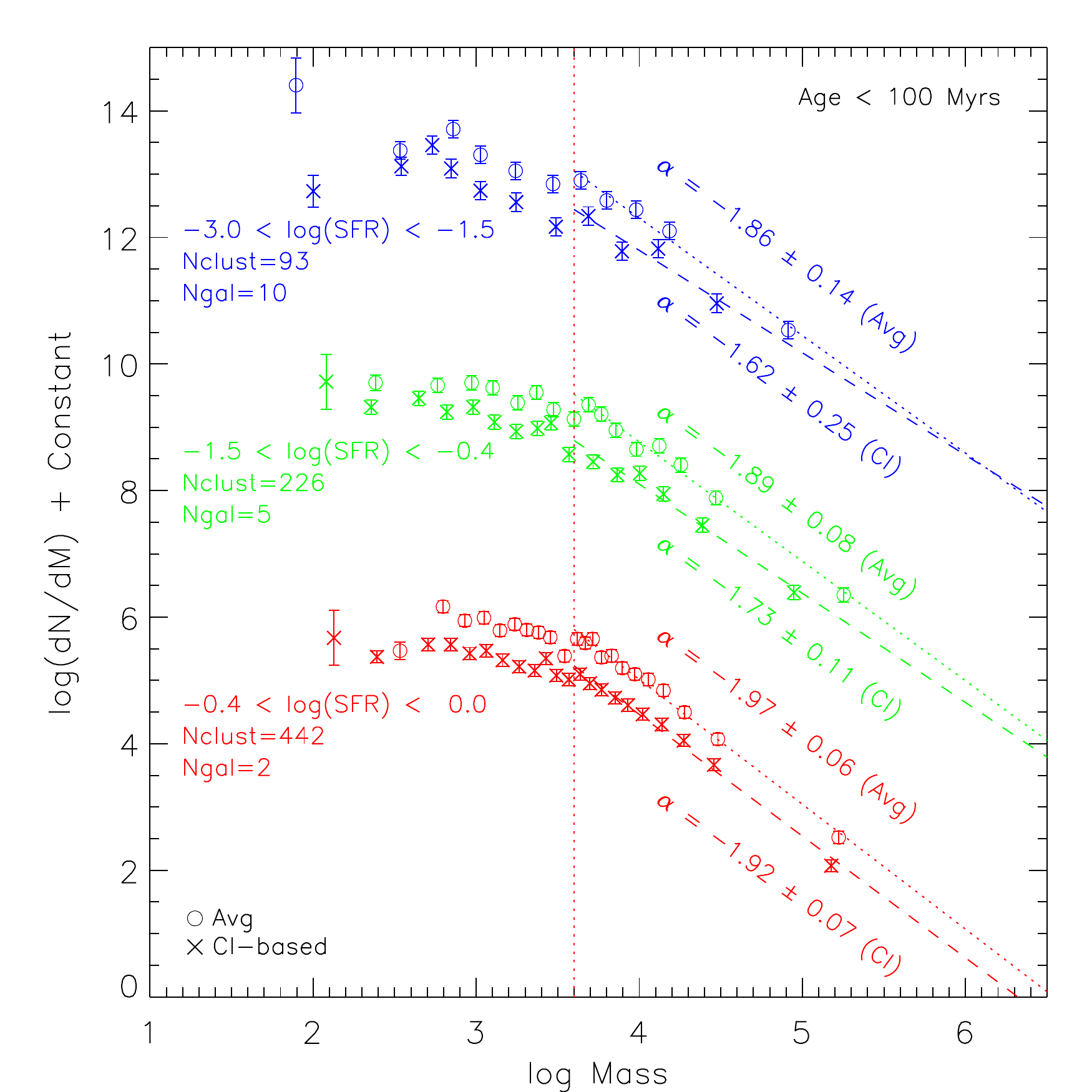}
  \caption{The mass functions (MFs) for the clusters in all LEGUS dwarf galaxies binned by global SFR. A single limiting log(mass) of 3.7 is used to fit the power-laws for all three SFR bins (see Figure~\ref{fig:MFci}). We find no trend between the MF slope and the binned SFR. }
   \label{fig:binMF}
   \end{center}
\end{figure}  


Figure~\ref{fig:binMF} shows the MFs of all young clusters (age$<$100~Myr) in the LEGUS dwarfs binned by their host-galaxy's SFR. Similar to our findings for the LFs, we find no statistical difference between the MF slopes across the SFR bins where the median MF slope is --1.9$\pm$0.1. We find no differences in the MF slopes when using various SFR bin definitions nor a color cut to approximate a 100~Myr age cut. We also note that we find no trend between the LF and MF slopes when using higher age cuts (up to 1 Gyr) to increase the cluster number statistics.



\begin{figure}
  \includegraphics[scale=0.5]{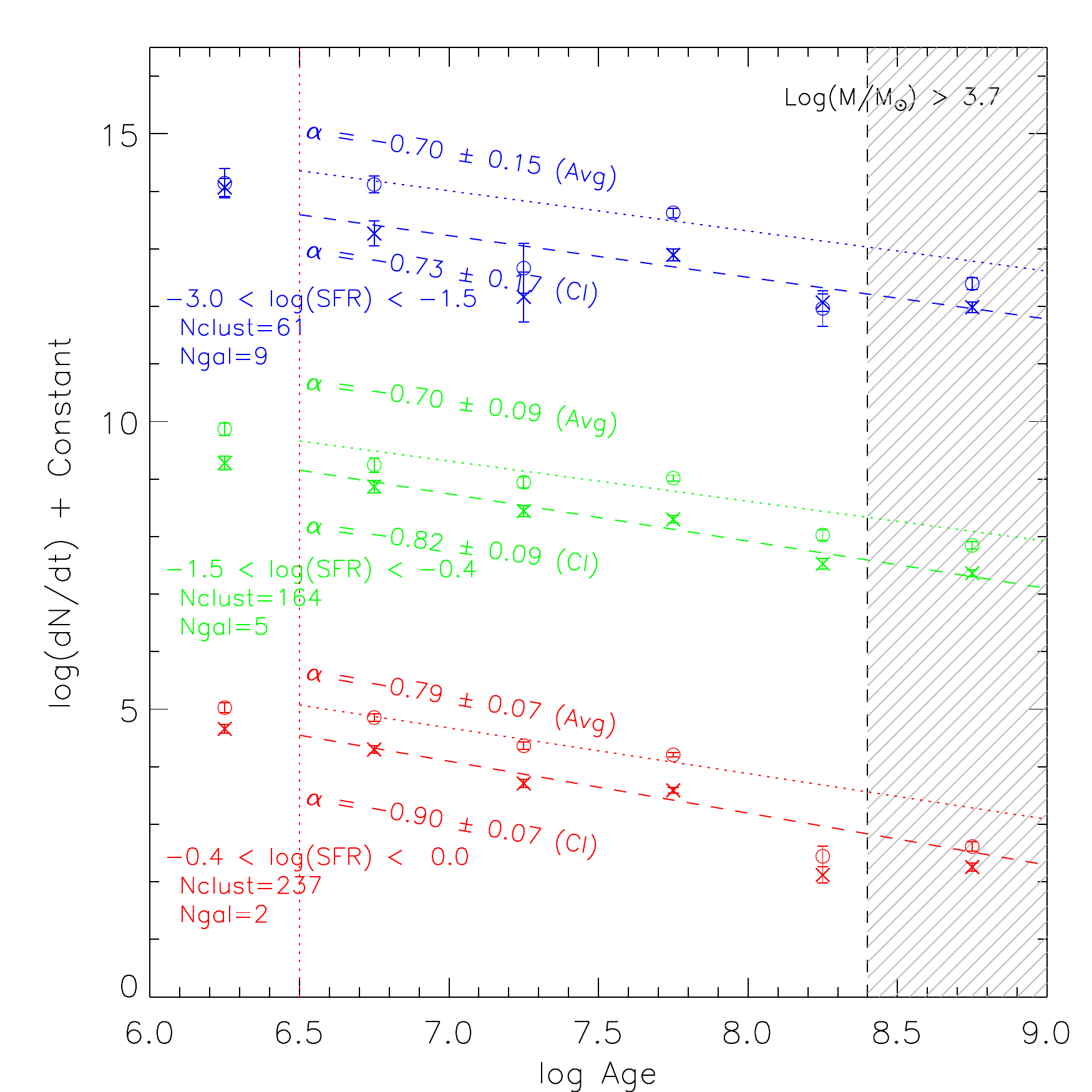}
  \caption{The age distributions for the clusters in all LEGUS dwarf galaxies binned by global SFR with a mass cut above log(mass) of 3.7. Following \citet{adamo17} and \citet{messa18a}, we exclude the youngest age bin in the fits (although the results are similar if this age bin is included). We find a constant slope of $-0.8\pm0.15$ and no trend in the age distributions across SFR bins.}
  \label{fig:bindndt}
\end{figure} 

Figure~\ref{fig:bindndt} shows the age distributions of all clusters binned by global SFR with a log(mass)$>$3.7 in the LEGUS dwarfs. We find a constant slope of $-0.8\pm0.15$ and no trend in the age distributions across SFR bins. We find similar results when using various age bin sizes, and we find slightly steeper slopes of $-1.0\pm0.1$ across the SFR bins when fitting to age bins above 10~Myrs. Additionally, if use a more conservative completeness mass cut of log(M/M$_{\odot}$)=4, we find similar results with a slightly flatter average slope of --0.75$\pm$0.15. 

The lack of any trends in the LFs, MFs, and age distributions across SFR bins suggests that clusters in different SFR environments exhibit similar mass (and similarly luminosity) distributions and similar disruption rates over time. We discuss this topic further in \S\ref{sec:disc}. We also note that an upcoming LEGUS paper will explore the luminosity, mass, and age distributions across different local environment in NGC4449 (See Whitmore in prep).




\subsubsection{MF Truncation} \label{sec:mftrunc}

We use two different methods to test whether or not there is a truncation in the composite mass function of the LEGUS dwarfs at the high end. Following Mok et al. (2018), we use the mass function in three different intervals of age: $1-10$, $10-100$, and $100-400$~Myr, where we apply a log(M) completeness cut of 3.4, 3.7, and 3.7 following the turnover in the MFs for each age interval (see Figure~\ref{fig:MFci}).

\begin{figure*}
  \includegraphics[scale=0.25]{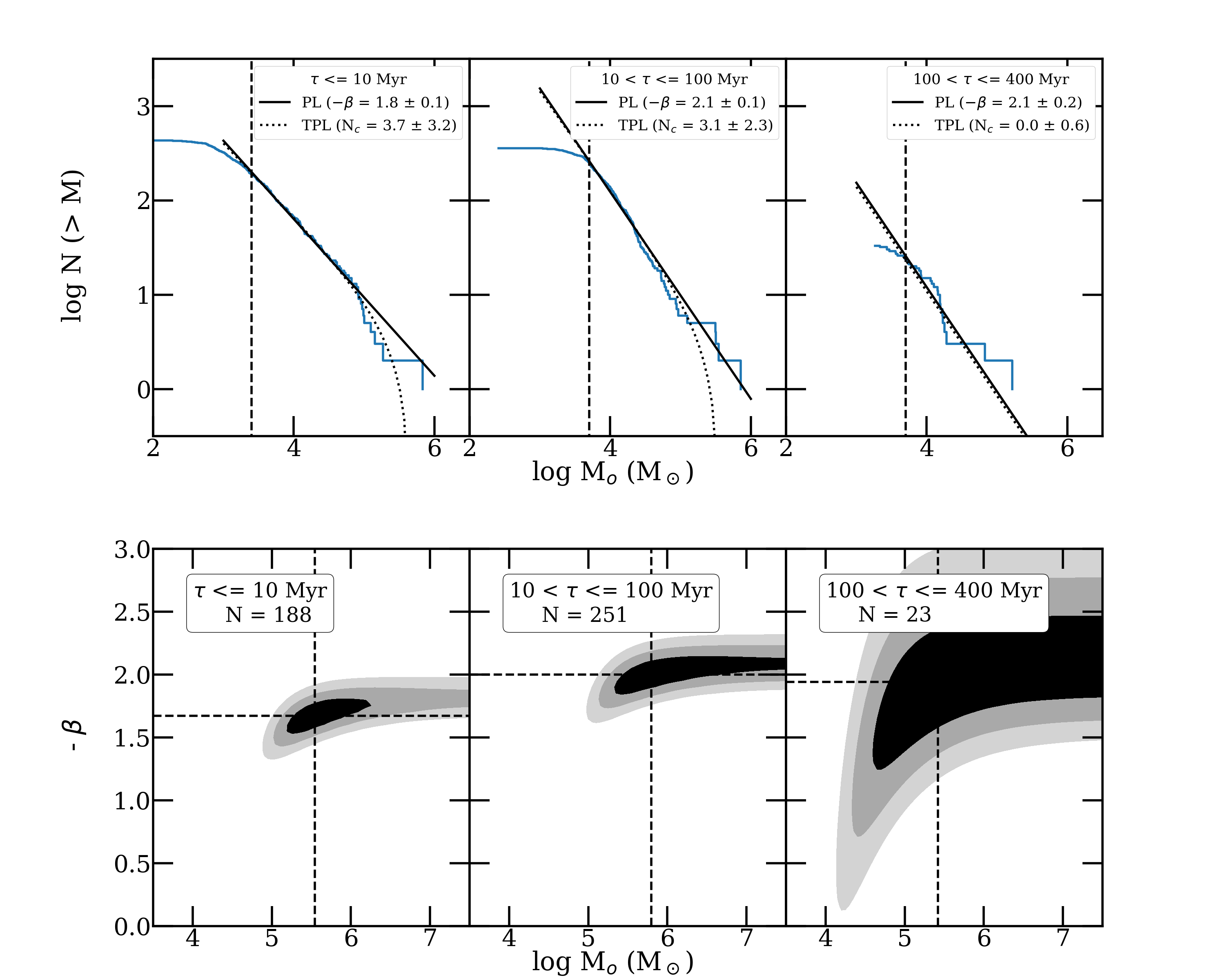}
  \caption{The figure shows the results from two independent methods used to test for a truncation at the upper end of the mass function for our composite LEGUS dwarf cluster sample.  The top panels show the results from the power law (dashed line) and truncated power law (dotted line) fits to the cumulative mass distributions in the $<10$~Myr (left), $10-100$~Myr (middle), and the $100-400$~Myr (right) intervals of age.  The log of best fit truncation mass ($M_0$) and the significance of this result ($N_0$) are also given in each panel, and show that this method does not find a statistically significant truncation mass.  The bottom panels show the 1-, 2-, and 3-$\sigma$ contours in the $\beta-M_0$ parameter space from a maximum likelihood fit to the masses, when an underlying Schechter function is assumed.  The contours do not close on any particular value of $M_0$, indicating that there is no statistically significant upper mass cutoff, just as found from the truncated power law fits.  The contours rule out upper mass cutoffs of $\approx10^5~M_{\odot}$ and below, but cannot rule out upper mass cutoffs higher than this.}
   \label{fig:mftrunc}
\end{figure*}

For method 1, we follow \citet{messa18a} and fit a truncated power law distribution to the cumulative mass distributions using the \textsc{mspecfit} software \citep{rosolowsky2005}. The best fit cutoff mass is characterized by $M_0$, and the significance of the fit can be determined from the accompanying value of $N_0$. The best fit results for clusters in the $1-10$, $10-100$, and $100-400$~Myr intervals are shown in the upper panels of Figure~\ref{fig:mftrunc}.  Here, the two youngest intervals return values for Log~$M_0\sim5.5-5.7$, but with low significance (only $\approx1\sigma$). The oldest $100-400$~Myr age range has too few clusters above the completeness limit to give a meaningful fit.  

For method 2, we perform a standard maximum likelihood analysis \citep[as described in Chapter 15.2 of][]{mo10} by assuming that the cluster masses have an underlying Schechter form.  This method returns the best fit values of the characteristic mass $M_0$ and power-law index $\beta$. We plot the resulting 1-, 2-, and 3-$\sigma$ confidence contours in the bottom panels of Figure~\ref{fig:mftrunc} for each of the three age intervals. The 2 younger age intervals do not show statistically significant evidence for an upper mass cutoff in the composite LEGUS dwarf sample since the 2 and 3$\sigma$ contours do not close (i.e., remain open to the right edge of each diagram). The oldest age interval contains too few clusters (N=23) for a robust measurement, and clearly demonstrates that the size of the contours is a strong function of the number of clusters in the sample when compared to the younger 2 age intervals. We find similar results when the larger age intervals of 1-200 and 1-400 Myr are used.

Given the shape of the 3$\sigma$ contours at the low mass end in the 2 younger age intervals, we can rule out a truncation mass of $\approx10^5~$M$_{\odot}$ and below, but cannot rule out that a truncation exists above this. We note that we find similar results in the 3 age ranges when removing the 3 highest SFR dwarfs (NGC3738, NGC4656, and NGC4449), except that the lower limit on the mass truncation is smaller at $\approx$10$^{4.5}~M_{\odot}$ in the 1-10~Myr age range. 

The overall results from these two independent methods are similar: neither one finds statistically significant evidence for a truncation at the upper end of the cluster mass function of young clusters in our composite LEGUS dwarf sample.  In addition, the maximum likelihood results indicate that any truncation mass must be higher than $\approx10^{4.5}~M_{\odot}$.

\section{DISCUSSION }\label{sec:disc}

The luminosity, mass, and age distributions of the star clusters in dwarf galaxies provide important clues to their formation and disruption. In this section, we discuss our results in this context. 



In \S\ref{sec:distfunc} and \S\ref{sec:trends}, we found that the luminosity functions of clusters in the LEGUS dwarf galaxies can be described by a simple power-law, $dN/dL\propto L^{\alpha}$ with $\alpha\approx-2$. This is similar to the cluster populations in two LEGUS spiral galaxies, which have higher SFRs.  NGC628 has a SFR$_{FUV+24\mu m}$=6.8~M$_{\odot}\rm{yr}^{-1}$ and M51 has a SFR$_{FUV+24\mu m}$=2.9~M$_{\odot}\rm{yr}^{-1}$ \citep{lee09b,cook14c}. The luminosity functions for the clusters in these galaxies were derived using the same methodology as used here, resulting in a slope of --2.09$\pm$0.02 \citep{adamo17} and --2.02$\pm$0.03 \citep{messa18a} for NGC628 and M51, respectively. Both the spiral and dwarf galaxy luminosity function slopes are consistent with each other, and show no evidence of a trend between luminosity slope and galaxy SFR. 

We also found that the cluster mass functions in the LEGUS dwarf galaxies can be described by a single power-law with an index of $\beta\approx-2$ in different age intervals for clusters with ages up to $\approx400$~Myr. A consistent MF slope over different age ranges can provide clues into the disruption of clusters over time. We see no evidence for flattening at the low end of the cluster mass functions (above the completeness limits), which means that mass-dependent disruption (i.e., where lower mass clusters disrupt faster than higher mass ones), does not have a strong impact on the observable mass and age ranges of our cluster population. We also do not find a correlation between the power-law indices of our composite cluster mass distributions with the overall SFR of the host galaxies; although the masses of the most massive clusters increase with SFR as expected from sampling statistics. 

\begin{figure*}
  \includegraphics[scale=0.23]{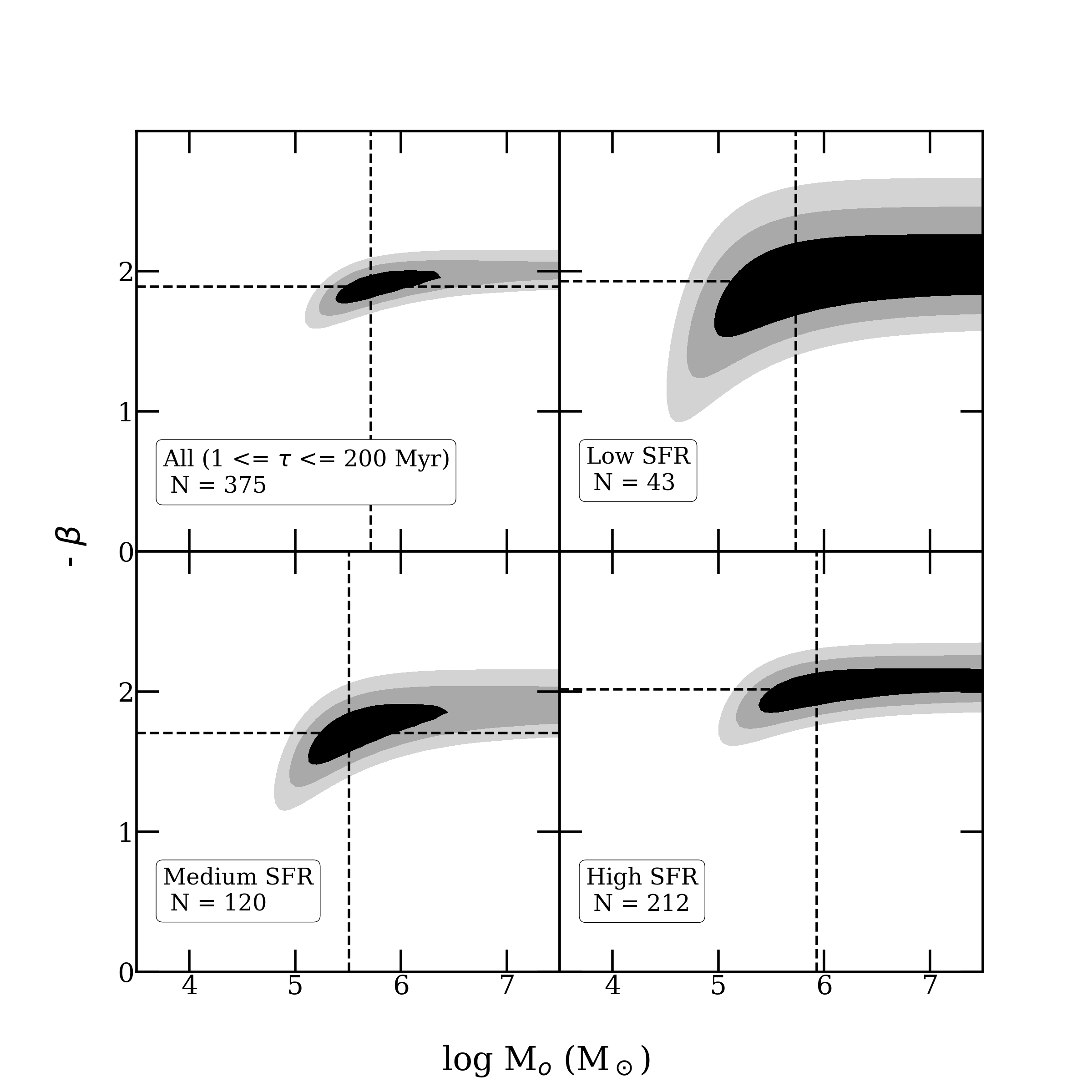}
  \caption{The maximum likelihood fits to the LEGUS cluster masses with an age of 1-200~Myr, where we show the 1-, 2-, and 3-$\sigma$ contours in the $\beta-M_0$ parameter space when an underlying Schechter mass function is assumed. The upper-left panel shows all clusters with an age of 1-200~Myr, while the remaining panels show the fits to clusters binned by global SFR using the same bins as in Figures~\ref{fig:binLF}--\ref{fig:bindndt}. We find no statistically significant evidence for a truncation mass at the 2-3$\sigma$ level due to the open contours, and that the size of the contours scales with the number of clusters in the sample. }
  \label{fig:MFtruncSFRbin}
\end{figure*}

Several previous studies have found truncations at the upper end of the cluster mass function for individual spiral and interacting galaxies \citep[$10^4<M_0~(M_{\odot})<10^6$;][]{gieles06b,jordan07,larsen09,bastian12a,adamo15,johnson17,adamo17,messa18a}. Most of the dwarf galaxies in our sample have fairly low SFRs and contain very few clusters, making it difficult to statistically test for a truncation in an individual galaxy. Therefore, in \S\ref{sec:mftrunc}, we tested a composite dwarf galaxy cluster mass function for a Schechter-like downturn at the high mass end. Two different methods found little evidence for an upper mass cutoff in two out of three age ranges (the third $100-400$~Myr range has too few clusters for a robust measurement). To put these results into context and to provide a more direct comparison to other LEGUS studies in the spirals M51 and NGC628 \citep{adamo17,messa18a} we test the age interval of 1-200~Myrs, which also provides the added benefit of better clusters statistics. The upper-left panel of Figure~\ref{fig:MFtruncSFRbin} shows the maximum likelihood contours for our composite dwarf sample in the age interval of 1-200~Myrs. We find no statistically significant evidence for a downturn at the 2-3$\sigma$ level (i.e., the contours remain open). 

In addition, we test if our MF truncation constraints change with global SFR in the remaining 3 panels of Figure~\ref{fig:MFtruncSFRbin}. Here, we bin the clusters with ages of 1-200~Myrs into the same SFR bin definitions as used in our LF, MF, and age distribution tests of Figures~\ref{fig:binLF}--\ref{fig:bindndt}. We find no significant evidence for an upper mass truncation at the 2-3$\sigma$ level in any SFR bin. We also find that the low-mass end of the 3$\sigma$ contours in the lowest SFR bin extends to lower truncation masses when compared to the higher SFR bins. However, since the number of clusters in bins of SFR scales with SFR and the size of the contours, this may give the appearance of trends in parameter constraints with SFR. To draw definitive conclusions, the size of the sample in the lowest SFR bin must be increase by a factor of 3 to 5 in future studies. Taking into account all of our MF truncation tests using different age intervals and SFR binned samples, the 1$\sigma$ contours are closed for some of the age intervals and SFR bins, while the 2-3$\sigma$ contours do not close in any of our tests. This indicates weak evidence ($<2\sigma$ level) for a truncation in some cases.


In Section~\ref{sec:trends}, we found that the age distributions of the clusters decline steadily, and can be described as a simple power law, $dN/d\tau \propto \tau^{\gamma}$, with $\gamma=-0.8\pm0.15$. The declining shape of the age distribution in the LEGUS dwarfs is remarkably robust to binning, mass range, age range of the fit, and the specific galaxies that are included.

In order to interpret this result, we need to disentangle the effects of formation versus disruption, since the observed distribution includes both the formation and disruption histories of the clusters: $\gamma_{cl} = \gamma_{\rm form} + \gamma_{\rm disrupt}$. We can do this by assuming that the cluster formation history is proportional to the star formation history, and estimating a composite formation history by summing the SFRs in different age ranges (i.e., the star formation histories; SFHs). As we are using a composite dwarf galaxy SFH from many independent systems, then presumably the combined SFH should be relatively flat over the past few hundred million years since bursts that occur in any individual galaxy should be uncorrelated. 

To test this assumption we utilize SFRs from two independent methods: 1) the H$\alpha$ and FUV SFRs from integrated light measurements corrected for internal dust extinction \citep{lee09b} and 2) the recent star formation histories from resolved-star CMD analysis \citep{cignoni18}. The integrated light measurements provide a low-resolution SFH since the SFRs derived from H$\alpha$ and FUV probe $t<10$~Myr and $t<100$~Myr timescales, respectively \citep{kennicutt98}. The recent SFHs for 3 of the LEGUS dwarfs were presented by \citet{cignoni18}, and the others will be presented in an upcoming paper (Cignoni et al. in preparation).  While we wait for the final SFHs to become available for all of our galaxies, we can still assess whether the composite SFH is flat, declines, or increases for the 17 dwarf galaxies using preliminary SFHs. 

\begin{figure}
  \includegraphics[scale=0.5]{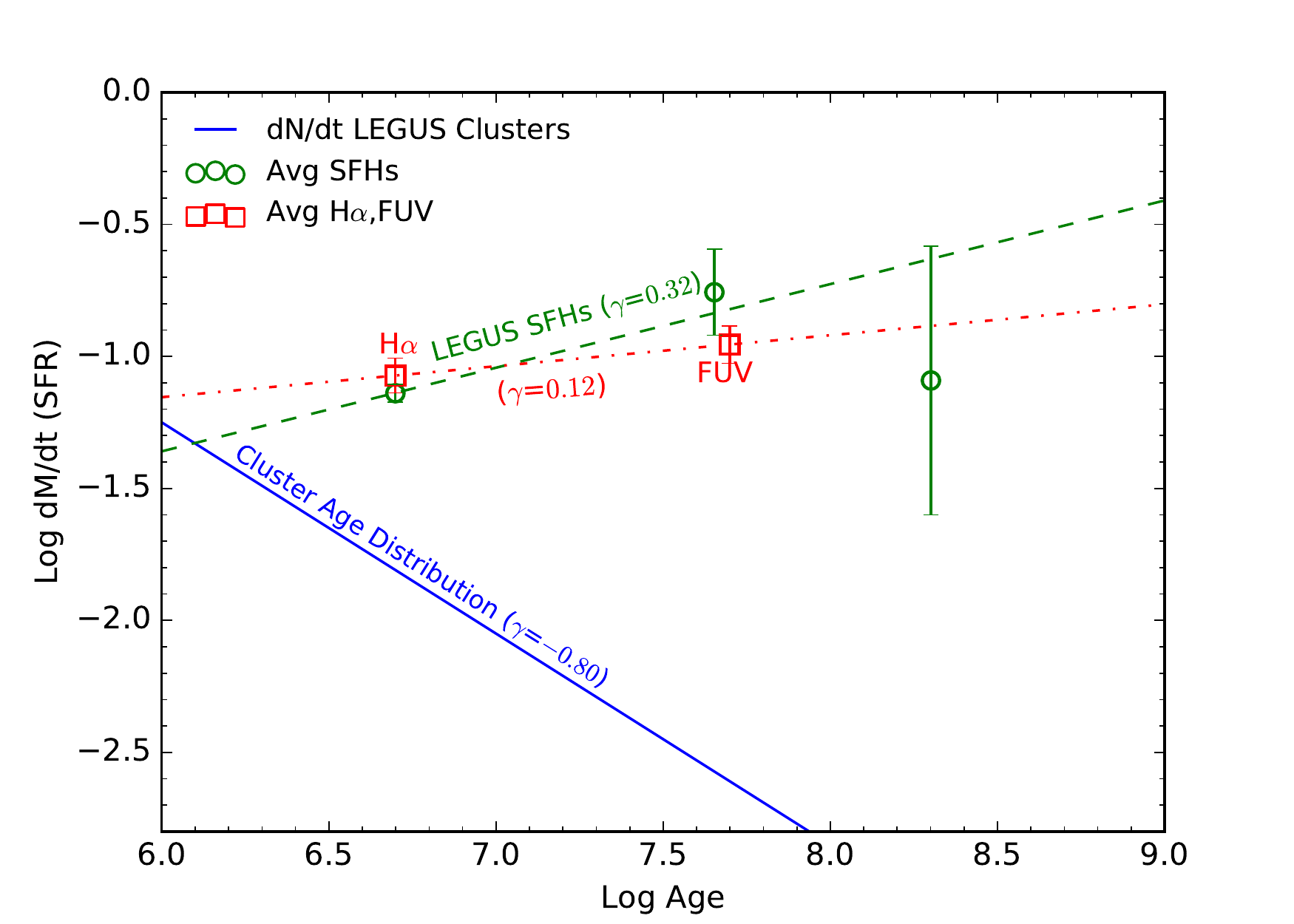}
  \caption{The total SFR versus age of the LEGUS dwarf galaxies using two independent SFR measurements. The red squares represent the summed H$\alpha$ and FUV SFRs corrected for dust extinction from \citet{lee09b}; we have updated the SFR conversion using the prescription of \citet{murphy11} with a Kroupa IMF. The green circles represent the summed SFRs derived from the resolved-star SFHs of \citet{cignoni18}. The SFHs are preliminary and are in the process of being updated. The blue solid line represents the cluster age distribution in the LEGUS dwarf sample that has been scaled to fit on this graph for purposes of comparing the slopes. The total SFRs from both methods show a constant star formation in the composite dwarf sample indicating that the decline in the cluster age distribution is dominated by cluster disruption. }
   \label{fig:superSFH}
\end{figure}

Figure~\ref{fig:superSFH} shows the average SFRs for the 17 LEGUS dwarfs over time using both the integrated light measurement and the resolved-star SFHs. We fit a power law to these SFRs (i.e., dM/dt vs t) and find a slope ($\gamma_{\rm form}$) in the range of 0.1--0.3, which is consistent with a flat or constant formation history over the age ranges studied here. This is similar to the results found by \citet{mcquinn10} for 18 nearby dwarf galaxies also using multi-band $HST$ observations. Since $\gamma_{\rm form} \approx 0$, then $\gamma_{\rm cl} \approx \gamma_{\rm disrupt}$ which means that the observed cluster age distribution is dominated by the disruption of clusters rather than their formation.  

The best fit power law ($\gamma$=--0.8$\pm$0.15) found in the LEGUS dwarfs, when compared with the SFHs and binned by different parameters, indicates that approximately 70-90\% of the clusters disrupt every decade in age, independent of cluster mass and SFR environment. These age distributions are similar to that found individually for a number of more massive galaxies \citep{fall05,villaLarsen11,fall12,silvavilla14,mulia16}, where a median $\gamma$ calculated in \citet{chandar17} is --0.7$\pm$0.3. This is consistent with the `quasi-universal' model of cluster formation and disruption \citep{whitmore03,whitmore07,fall12}. However, several studies have found age distributions significantly flatter ($\gamma$=--0.2 to --0.4) than that found in our dwarfs \citep{villaLarsen11,johnson17,adamo17,messa18a}, and several works have found evidence that cluster disruption may occur at different rates across different environments within the same galaxy \citep{bastian12a,silvavilla14,adamo17,messa18b}. For instance, \citet{messa18b} found a trend between the cluster disruption rate and gas surface density variations across the M51 disk. In a future work, we will investigate whether other parameters (e.g., SFR/area~$\equiv\Sigma_{\rm SFR}$) may have a more pronounced effect on the formation and disruption of clusters in our dwarf galaxies.

\section{SUMMARY}

This study has uniformly identified and examined the star clusters in a large sample of dwarf galaxies (N=17) with high resolution $HST$ imaging in 5 filters. The nearly uniform data has facilitated: 1) a detailed comparison of different cluster identification and photometry methods commonly used in the literature, 2) an examination of cluster properties in low-SFR environments with better number statistics than previously studied. The main conclusions are listed below. 

\begin{itemize} 

\item An examination of two widely used aperture corrections (average-based and concentration index (CI)-based) shows that both methods provide largely consistent colors and ages, but that roughly half of the clusters show CI-based aperture corrections that are inconsistent with the average correction given the measured errors. The median total flux difference derived from the two aperture corrections is 0.2~mag suggesting that many of the clusters are more compact than the average training cluster. This median total flux difference translates into a mass offset of 0.1--0.2~dex between the two aperture correction methods. However, the ensemble luminosity, mass, and age distributions derived from both aperture corrections are consistent with each other within the errors. 

\item Comparing the LEGUS cluster catalog with that of a previous large sample of dwarf galaxies \citep{cook12} shows that the LEGUS catalog is more complete and provides more accurate total fluxes. The differences in the total fluxes is attributed to the low resolution of the ground-based imaging used to augment the HST imaging in \citet{cook12}. For clusters found in common to both catalogs, we find overall agreement in the ages, but the \citet{cook12} masses can be considerably different given the large total flux differences.

\item The luminosity and mass functions observed for clusters in the LEGUS dwarfs can be described by a power-law, with an index of $\approx-2$.  The mass function appears to be independent of cluster age up to the $\approx400$~Myr studied here, and does not vary with star formation rate.

\item The composite cluster mass function shows little evidence for an upper mass truncation at the 2-3$\sigma$ level. The lack of a significant evidence holds for different age intervals and cluster samples binned by global SFR. The extent of the 3$\sigma$ contours in the maximum likelihood fits rule out a truncation below $\approx$10$^{4.5}$~M$_{\odot}$, but cannot rule out a truncation at higher masses.

\item The observed age distribution for the composite cluster population in the LEGUS dwarf galaxies can be described by a power law, $dN/d\tau \propto \tau^{\gamma}$, with $\gamma = -0.8\pm0.15$, over the age range $\approx10-400$~Myr.  This distribution appears to be independent of the mass of the clusters, and does not vary with star formation rate. 

\item  The composite star formation history for our dwarf galaxies from both integrated light measurements (H$\alpha$ and FUV) and preliminary resolved-star CMD SFHs are both quite flat, with a best fit power law index of $\gamma_{\rm form}=0.1-0.3$.  This indicates that disruption dominates the observed cluster age distribution, with $\approx$80\% of the clusters being disrupted every decade in age.

\end{itemize} 

In a future work, we will use updated star formation rates for the LEGUS dwarf galaxies to determine the star formation rate density, $\Sigma_{\rm SFR}$ in a consistent way.  We will also extrapolate the mass functions presented here to determine the fraction of stars found in clusters ($\Gamma$) in the LEGUS dwarf galaxies, and compare the results with the more massive galaxies in the LEGUS sample. Finally, we will explore if the age distributions change as a function of $\Sigma_{\rm SFR}$.



\section*{Acknowledgements}

We thank the referee for their helpful comments. Based on observations made with the NASA/ESA Hubble Space Telescope, obtained at the Space Telescope Science Institute, which is operated by the Association of Universities for Research in Astronomy, Inc., under NASA contract NAS 5-26555. These observations are associated with program \# 13364. This research has made use of the NASA/IPAC Extragalactic Database (NED) which is operated by the Jet Propulsion Laboratory, California Institute of Technology, under contract with NASA.  A.A. acknowledges the support of the Swedish Research Council (Vetenskapsradet) and the Swedish National Space Board (SNSB). D.A.G acknowledges support by the German Aerospace Center (DLR) and the Federal Ministry for Economic Affairs and Energy (BMWi) through program 50OR1801 ''MYSST: Mapping Young Stars in Space and Time''. 

\bibliographystyle{tex/mn2e}   
\bibliography{tex/all}

\newpage

\end{document}

%% file: tables/GlobalProp.tex
\begin{table*}
{Global Properties of the LEGUS Dwarf Galaxies}\\
\begin{tabular}{lcccccccc}
\hline
\hline
Galaxy         & RA	       & DEC	       & T	       & $D$   & 12+Log(O/H)   & Method        & SFR (FUV+24$\mu m$)	       & Mstar         \\ 
Name	       & (J2000.0)     & (J2000.0)     &               & (Mpc) &	       &	       & ($M_{\odot}$yr$^{-1}$)& ($M_{\odot}$) \\ 
(1)	       & (2)	       & (3)	       & (4)	       & (5)   & (6)	       & (7)	       & (8)		       & (9)	     \\ 
\hline
UGC685 & 01:07:25.8 & +16:41:35.5 & 9 & 4.83 & 8.00 $\pm$ 0.03 & 4363\AA & 3.85e-03 & 8.82e+07 \\
UGC695 & 01:07:46.4 & +01:03:49.2 & 6 & 10.90 & 7.69 $\pm$ 0.12 & 4363\AA & 9.41e-03 & 1.63e+08 \\
UGC1249 & 01:47:29.9 & +27:20:00.0 & 9 & 6.90 & 8.62 $\pm$ 0.20 & Strong & 9.38e-02 & 9.76e+08 \\
NGC1705 & 04:54:13.7 & -53:21:40.9 & 11 & 5.10 & 8.21 $\pm$ 0.05 & 4363\AA & 4.84e-02 & 2.44e+08 \\
ESO486-G021$^{\dagger}$ & 05:03:19.6 & -25:25:22.6 & 2 & 9.50 & $\ldots$ & $\ldots$ & 2.54e-02 & 1.82e+08 \\
UGC4305 & 08:19:13.8 & +70:42:43.1 & 10 & 3.05 & 7.92 $\pm$ 0.10 & 4363\AA & 5.54e-02 & 3.13e+08 \\
UGC4459 & 08:34:05.0 & +66:10:59.2 & 10 & 3.66 & 7.82 $\pm$ 0.09 & 4363\AA & 4.38e-03 & 1.90e+07 \\
UGC5139 & 09:40:31.5 & +71:11:23.2 & 10 & 3.98 & 7.92 $\pm$ 0.05 & 4363\AA & 7.37e-03 & 4.98e+07 \\
IC559 & 09:44:43.8 & +09:36:54.0 & 5 & 5.30 & 8.07 $\pm$ 0.10 & 4363\AA & 2.77e-03 & 6.92e+07 \\
UGC5340$^{\dagger}$ & 09:56:46.2 & +28:49:15.3 & 10 & 12.70 & 7.20 $\pm$ 0.05 & 4363\AA & 1.55e-02 & 3.71e+07 \\
NGC3274$^{\dagger}$ & 10:32:17.3 & +27:40:08.0 & 7 & 6.55 & 8.01 $\pm$ 0.20 & Strong & 3.89e-02 & 3.23e+08 \\
NGC3738 & 11:35:47.1 & +54:31:32.0 & 10 & 4.90 & 8.04 $\pm$ 0.06 & 4363\AA & 3.63e-02 & 5.12e+08 \\
UGC7242$^{\dagger}$ & 12:14:09.2 & +66:04:45.1 & 6 & 5.42 & $\ldots$ & $\ldots$ & 5.67e-03 & $\ldots$ \\
NGC4248$^{\dagger}$ & 12:17:49.9 & +47:24:33.1 & 3 & 7.80 & 8.15 $\pm$ 0.20 & Strong & 8.39e-03 & 7.71e+08 \\
UGC7408 & 12:21:15.2 & +45:48:43.0 & 9 & 6.70 & $\ldots$ & $\ldots$ & 1.13e-02 & 2.19e+08 \\
UGCA281 & 12:26:15.8 & +48:29:38.8 & 11 & 5.90 & 7.80 $\pm$ 0.03 & 4363\AA & 1.54e-02 & 4.18e+07 \\
NGC4449 & 12:28:12.0 & +44:05:42.3 & 10 & 4.31 & 8.32 $\pm$ 0.03 & 4363\AA & 4.53e-01 & 3.01e+09 \\
NGC4485$^{\dagger}$ & 12:30:31.9 & +41:41:33.8 & 10 & 7.60 & $\ldots$ & $\ldots$ & 9.26e-02 & 7.59e+08 \\
NGC4656 & 12:43:57.5 & +32:10:13.3 & 9 & 5.50 & 8.09 $\pm$ 0.05 & 4363\AA & 7.58e-01 & 2.48e+09 \\
IC4247 & 13:26:43.4 & -30:21:40.8 & 2 & 5.11 & 8.27 $\pm$ 0.20 & Strong & 4.11e-03 & 5.48e+07 \\
NGC5238 & 13:34:45.0 & +51:35:57.4 & 8 & 4.51 & 7.96 $\pm$ 0.20 & Strong & 9.77e-03 & 1.17e+08 \\
NGC5253 & 13:39:55.9 & -31:38:24.0 & 11 & 3.15 & 8.15 $\pm$ 0.10 & 4363\AA & 2.73e-01 & 8.73e+08 \\
NGC5477 & 14:05:33.3 & +54:27:39.0 & 9 & 6.40 & 7.95 $\pm$ 0.02 & 4363\AA & 2.76e-02 & 1.55e+08 \\
\hline
\end{tabular}

\input{tables/GlobalProp.txt}
\label{tab:genprop}

\end{table*}

%% file: tables/GlobalProp.txt
\caption{Properties of the LEGUS dwarf galaxies. Column 1: Galaxy name. Column 2 and 3: J2000 right ascension and declination from NED. Column 4: Galaxy morphological type from \citet{legus}.  Column 5: Distance in Mpc from \citet{legus}.  Column 6: The gas-phase metallicity and error as compiled by \citet{cook14c}. Column 7: The method used to calculate the metallicity in Column 6. Column 8: The SFR based on the FUV fluxes of \citet{lee11} inside the IR-based apertures of \citet{dale09} and corrected for internal dust extinction via the \citet{hao11} prescription.  Column 9:  The stellar mass as computed by \citet{cook14c}, which is derived from the Spitzer IRAC channel 1 (3.6$\mu m$) filter. The $^{\dagger}$ symbol represents the dwarf galaxies whose cluster catalogs are not finalized. }

%% file: tables/ClustProp.tex
\begin{table}

\centering{Star Cluster Statistics in the LEGUS Dwarf Galaxies} \\ 
\begin{tabular}{ccccc}
\hline
\hline
 Galaxy	&N 		&N        &N         &N         \\ 
      	&Candidates	&Class 1,2&  Class 3 &Class 4   \\ 
(1)	& (2)		& (3)	  & (4)	     & (5)      \\ 
\hline

UGC685 & 20 & 11 & 3 & 6 \\
UGC695 & 111 & 11 & 6 & 94 \\
UGC1249 & 220 & 48 & 40 & 132 \\
NGC1705 & 96 & 29 & 13 & 54 \\
UGC4305 & 199 & 33 & 27 & 139 \\
UGC4459 & 30 & 7 & 3 & 20 \\
UGC5139 & 39 & 9 & 7 & 23 \\
IC559 & 43 & 21 & 4 & 18 \\
NGC3738 & 435 & 141 & 86 & 208 \\
UGC7408 & 69 & 34 & 11 & 24 \\
UGCA281 & 49 & 11 & 4 & 34 \\
NGC4449 & 1367 & 321 & 177 & 869 \\
NGC4656 & 431 & 184 & 78 & 169 \\
IC4247 & 45 & 5 & 3 & 37 \\
NGC5238 & 18 & 8 & 1 & 9 \\
NGC5253 & 231 & 57 & 23 & 151 \\
NGC5477 & 72 & 14 & 9 & 49 \\
\hline 
Total & 3475 & 944 & 495 & 2036 \\

\hline
\end{tabular}
\input{tables/clustprop.txt}
\label{tab:clustprop}

\end{table}

%% file: tables/clustprop.txt
\caption{ Column 1: Galaxy name. Column 2: The total number of cluster candidates found by the LEGUS extraction tool. Column 3: The number of cluster candidates with a visual classification of 1 and 2.  Column 4: The number of cluster candidates with a visual classification of 3 (stellar associations). Column 5: The number of cluster candidates with a visual classification of 4 (contaminants). Column 6: The total number of human-identified clusters with a visual classification of 1, 2, and 3. }

%% file: tables/ApCorr.tex
\begin{table*}
\centering{Average Aperture Corrections}\\
\begin{tabular}{cccccccccccc}

\hline
\hline
 Galaxy	&F275W	&F336W	&F438W	&F555W	&F814W	&F435W	&F555W	&F606W	&F814W	&Range 	&N Training 	\\ 
Name	&WFC3	&WFC3	&WFC3	&WFC3	&WFC3	&ACS	&ACS	&ACS	&ACS	&ApCorr	&Clusters	\\ 
(1)	& (2)	& (3)	& (4)	& (5)	& (6)	& (7)	& (8)	& (9)	& (10)	& (11)	& (12)		\\ 
\hline

eso486g021 & -0.98 & -1.08 & -0.96 & -0.83 & -0.86 & $\ldots$ & $\ldots$ & $\ldots$ & $\ldots$ & 0.25 & 7 \\
ic4247 & -0.94 & -0.80 & -1.08 & $\ldots$ & $\ldots$ & $\ldots$ & $\ldots$ & -0.68 & -0.80 & 0.40 & 2 \\
ic559 & -0.97 & -0.80 & -0.77 & -0.83 & -1.05 & $\ldots$ & $\ldots$ & $\ldots$ & $\ldots$ & 0.28 & 9 \\
ngc1705 & -1.07 & -0.86 & -0.91 & -0.91 & -1.01 & $\ldots$ & $\ldots$ & $\ldots$ & $\ldots$ & 0.21 & 12 \\
ngc3738 & -0.97 & -0.79 & -0.84 & $\ldots$ & $\ldots$ & $\ldots$ & $\ldots$ & -0.88 & -0.96 & 0.18 & 21 \\
ngc4449 & -0.91 & -0.90 & $\ldots$ & $\ldots$ & $\ldots$ & -0.85 & -0.87 & $\ldots$ & -0.97 & 0.12 & 55 \\
ngc4656 & -0.91 & -0.80 & -0.79 & -0.85 & -0.88 & $\ldots$ & $\ldots$ & $\ldots$ & $\ldots$ & 0.13 & 51 \\
ngc5238 & -1.03 & -0.97 & -0.78 & $\ldots$ & $\ldots$ & $\ldots$ & $\ldots$ & -0.86 & -0.86 & 0.25 & 9 \\
ngc5253 & -0.92 & -0.97 & $\ldots$ & $\ldots$ & $\ldots$ & -0.95 & -0.92 & $\ldots$ & -0.85 & 0.12 & 15 \\
ngc5477 & -1.10 & -1.05 & -0.98 & $\ldots$ & $\ldots$ & $\ldots$ & $\ldots$ & -0.86 & -0.95 & 0.25 & 5 \\
ugc1249 & -0.84 & -0.84 & -0.75 & $\ldots$ & $\ldots$ & $\ldots$ & $\ldots$ & -0.91 & -0.98 & 0.23 & 26 \\
ugc4305 & -0.85 & -0.87 & -0.81 & $\ldots$ & $\ldots$ & $\ldots$ & -0.75 & $\ldots$ & -0.87 & 0.12 & 5 \\
ugc4459 & -0.58 & -0.66 & -0.81 & $\ldots$ & $\ldots$ & $\ldots$ & -0.73 & $\ldots$ & -0.87 & 0.29 & 4 \\
ugc5139 & -0.79 & -0.59 & -0.69 & $\ldots$ & $\ldots$ & $\ldots$ & -0.70 & $\ldots$ & -0.84 & 0.25 & 6 \\
ugc5340 & -0.79 & -0.82 & -0.81 & $\ldots$ & $\ldots$ & $\ldots$ & $\ldots$ & -0.77 & -0.86 & 0.10 & 15 \\
ugc685 & -0.95 & -0.75 & -0.76 & $\ldots$ & $\ldots$ & $\ldots$ & $\ldots$ & -0.76 & -0.85 & 0.20 & 6 \\
ugc695 & -0.96 & -0.81 & -0.76 & -1.12 & -1.02 & $\ldots$ & $\ldots$ & $\ldots$ & $\ldots$ & 0.36 & 5 \\
ugc7242 & -1.13 & -0.81 & -0.80 & $\ldots$ & $\ldots$ & $\ldots$ & $\ldots$ & -0.90 & -0.81 & 0.33 & 2 \\
ugc7408 & -0.82 & -0.71 & -0.78 & $\ldots$ & $\ldots$ & $\ldots$ & $\ldots$ & -0.90 & -0.91 & 0.20 & 21 \\
ugca281 & -0.62 & -0.69 & -0.95 & $\ldots$ & $\ldots$ & $\ldots$ & $\ldots$ & -1.07 & -1.04 & 0.45 & 2 \\
ngc4485 & -0.76 & -0.75 & $\ldots$ & $\ldots$ & -0.84 & -0.79 & $\ldots$ & -0.77 & $\ldots$ & 0.09 & 22 \\
ngc3274 & -0.42 & -0.69 & -0.74 & -0.70 & -0.82 & $\ldots$ & $\ldots$ & $\ldots$ & $\ldots$ & 0.39 & 8 \\
ngc4248 & -0.94 & -0.90 & -0.79 & -0.82 & -0.88 & $\ldots$ & $\ldots$ & $\ldots$ & $\ldots$ & 0.15 & 8 \\
\hline
\end{tabular}

\input{tables/apcorr.txt}
\label{tab:apcorr}

\end{table*}

%% file: tables/apcorr.txt
\caption{The average aperture corrections for the LEGUS dwarfs across all filters, where different filter combinations exist for different galaxies. Column 1: Galaxy name. Columns 2-10: The average aperture correction for each filter. Column 11: The range (maximum minus minimum) average aperture correction across the filters for a galaxy. Column 12: The number of training clusters used to derive the average aperture correction.}

%% file: tables/fakeclust.tex
\begin{table*}
\centering{CI-based Aperture Corrections}\\
\begin{tabular}{cccccccc}

\hline
\hline
 Camera	&Aperture	&C0	&C1	&C2	&C3	& CI	& CI	\\ 
	&Radius	 	&	&	&	&	& min	& max	\\ 
        &(pixels)       & (\#)  & (\#)  & (\#)  & (\#)  & (mag) & (mag) \\ 
\hline

acs & 4 & -4.811 & 7.132 & -2.733 & -0.005 & 1.30 & 2.23 \\
wfc3 & 4 & -2.357 & 3.919 & -1.722 & -0.019 & 1.10 & 2.24 \\
acs & 5 & -4.452 & 6.464 & -2.347 & -0.045 & 1.30 & 2.23 \\
wfc3 & 5 & -2.220 & 3.633 & -1.532 & -0.029 & 1.10 & 2.24 \\
acs & 6 & -4.092 & 5.812 & -1.991 & -0.088 & 1.30 & 2.23 \\
wfc3 & 6 & -2.100 & 3.396 & -1.395 & -0.036 & 1.10 & 2.24 \\
\hline
\end{tabular}

\input{tables/fakeclust.txt}
\label{tab:fakeclust}

\end{table*}

%% file: tables/fakeclust.txt
\caption{The cubic polynomial fits between aperture correction and CI for both WFC3 and ACS cameras using the 3 photometric apertures allowed in the LEGUS extraction tool (4, 5, and 6 pixels). The last two columns represent the maximum measured fake star CI (CI min) and the maximum measured fake cluster CI (CI max); these two quantities represent the CI limits for which our CI-based aperture corrections are valid. The aperture correction are calculated via: correction$ = C0 + C1 \cdot CI + C2 \cdot CI^2 + C3 \cdot CI^3$}.

%% file: tables/clustnum.tex
\begin{table}
\centering{LEGUS-ANGST Star Cluster Comparison}\\
\begin{tabular}{ccccc}

\hline
\hline
Galaxy         & N     &  N     &  N           & N           \\ 
Name           & ANGST &  LEGUS &  ANGST       & LEGUS       \\ 
	       & All   &  All   &  $<$100~Myr  & $<$100~Myr \\ 
	       & (\#)  &  (\#)  &  (\#)        & (\#)        \\ 
\hline

All & 19 & 49 & 4 & 43 \\
UGC4305 & 10 & 33 & 3 & 30 \\
UGC4459 & 7 & 7 & 1 & 6 \\
UGC5139 & 2 & 9 & 0 & 7 \\
\hline
\end{tabular}

\input{tables/clustnum.txt}
\label{tab:clustnum}

\end{table}

%% file: tables/clustnum.txt
\caption{Comparison of cluster numbers between ANGST and LEGUS. The numbers LEGUS reflect those for class 1 and 2 clusters only; the ANGST goal was to only find gravitationally bound clusters so the apples to apples comparison is class 1 and 2. The LEGUS extraction tool finds 2.5 and 11 times the number of clusters found in ANGST for all ages and $<100$~Myr, respectively. }